\documentclass[10pt,amssymb,amsmath,nobibnotes,nofootinbib,aps,prb,showpacs,floatfix]{revtex4-2} 
\usepackage{graphicx}
\usepackage{multirow}
\usepackage{bm}
\usepackage{dcolumn}
\usepackage[dvipsnames]{xcolor}
\usepackage{graphicx}
\usepackage{dcolumn}
\usepackage{bm}
\usepackage{soul}
\usepackage{float}
\usepackage[hidelinks=true]{hyperref}
\usepackage[hyphenbreaks]{breakurl}
\hypersetup{
  colorlinks   = true, 
  urlcolor     = blue, 
  linkcolor    = blue, 
  citecolor    = blue 
}
\begin{document}
\renewcommand{\thefootnote}{\fnsymbol{footnote}}


\title{\texorpdfstring{Interplay of Orbital Degeneracy and Vacancies in Stabilizing Collinear Magnetic Order in  Cr$_{1+\delta}$Te$_2$}
{}}

\author{Prasanta Chowdhury$^{1,\ddagger}$, Jyotirmoy Sau$^{2,6,\ddagger}$, Mohamad Numan$^1$, Jhuma Sannigrahi$^3$, Matthias Gutmann$^4$, Gangadhar Das$^{5}$, D. T. Adroja$^{4,7}$, Saurav Giri$^1$, Manoranjan Kumar$^2$, Subham Majumdar$^1$}
\email{sspsm2@iacs.res.in}
\thanks{$^\ddagger$These authors contributed equally to this work.}

\affiliation{$^1$School of Physical Sciences, Indian Association for the Cultivation of Science, 2A \& B Raja S. C. Mullick Road, Jadavpur, Kolkata 700 032, India}

\affiliation{$^2$Department of Condensed Matter and Materials Physics, S. N. Bose National Centre for Basic Sciences, JD Block, Sector III, Salt Lake, Kolkata 700106, India}

\affiliation{$^3$School of Physical Sciences, Indian Institute of Technology Goa, Farmagudi, Goa 403401, India}

\affiliation{$^4$ISIS Neutron and Muon Source, Science and Technology Facilities Council, Rutherford Appleton Laboratory, Chilton Didcot OX11 0QX, United Kingdom}

\affiliation{$^5$Elettra-Sincrotrone Trieste, Strada Statale 14 km, km 163.5 in AREA Science Park 34149, Trieste, Italy}

\affiliation{$^6$Department of Physics and Astronomy, Uppsala University, Box-516, S-75120 Uppsala, Sweden}

\affiliation{$^7$Highly Correlated Matter Research Group, Physics Department, University of Johannesburg, Auckland Park 2006, South Africa}

\begin{abstract}
Cr$_{1+\delta}$Te$_2$, a two-dimensional van der Waals ferromagnet, displays a contested magnetic structure, poised between collinear and non-collinear spin configurations. In this work, we investigate the magnetic structure of Cr$_{1.33}$Te$_2$ at the microscopic level by combining single-crystal neutron diffraction, X-ray absorption spectroscopy, and first-principles calculations. Neutron diffraction measurements reveal a distinct collinear spin alignment, whereas spectroscopic analyses reveal inherent structural vacancies at both Cr and Te sites. These vacancies lead to local symmetry breaking that elevates the orbital degeneracy of the Cr 3$d$ states, as demonstrated by our first-principles analysis. The resulting modification of magnetocrystalline anisotropy emerges as the key mechanism stabilising the collinear magnetic ground state over the non-collinear one in the presence of vacancies. Our findings uncover a vacancy-driven route to control spin anisotropy and magnetic ordering in layered ferromagnets, offering new insights into the design of tunable 2D magnetic materials.

\end{abstract}

\maketitle

\noindent\textit{Introduction:} Vacancies play a decisive role in tuning the structural and magnetic properties of correlated materials, often inducing substantial changes in their magnetic ground states~\cite{MnCoGe, BaFeSe, LaMnSb2}, yet their microscopic influence remains only sparsely investigated. Here we study one member of the Cr$_{1+\delta}$Te$_2$ ($0 \leq \delta \leq 1$) family of binary van der Waals (vdW) compounds, which is inherently prone to vacancy formation and thus provides a natural platform for revealing vacancy-driven modifications of magnetic structure. These materials provide a fertile ground for exotic spin textures, such as skyrmions and other non-coplanar spin configurations, which manifest in distinctive transport signatures, including the topological Hall effect ({\sf THE})~\cite{Cr0.87Te_ACS_Nano, Cr1.33Te2_Rana_Saha, Cr1.53Te2_Adv, Cr5Te8_Nitesh, CrTe2_AdvF, Cr5Te8_AHE}. The intercalated Cr concentration, $\delta$, significantly influences the crystal structure, magnetic properties, and electronic transport characteristics of these compounds~\cite{Cr-Te_Phase, Cr1.53Te2_Adv, Fujisawa_Cr1+deltaTe2}. Studies have shown that as $\delta$ increases from 0.33 to 0.82, the magnetic Curie temperature ($T_C$) rises monotonically from 160~K to 350~K, accompanied by a transition in magnetic anisotropy from out-of-plane (OOP) to in-plane (IP) configurations~\cite{Cr1+deltaTe2_PRM, Cr1.53Te2_Adv}. This magnetic anisotropy plays a crucial role in stabilizing the magnetic configurations in these compounds.

\par
{\sf THE}, which arises from non-coplanar spin configurations with finite scalar spin chirality (SSC), has been observed in several Cr-Te-based compounds~\cite{Cr0.87Te_ACS_Nano, Cr1.53Te2_Adv, Cr5Te8_Nitesh, Cr1.2Te2_Nano_Letter, Cr2.76Te4, Cr1.61Te2_THE}. Recent studies using Lorentz Transmission Electron Microscopy (LTEM) have confirmed the presence of skyrmions in certain compositions, such as Cr$_{1.3}$Te$_2$~\cite{Cr1.33Te2_Rana_Saha, CrTe2_AdvF}, Cr$_5$Te$_8$~\cite{Cr5Te8_Nitesh} and Cr$_{1.53}$Te$_2$~\cite{Cr1.53Te2_Adv}. The existence of {\sf THE} and skyrmions suggests non-collinear spin configurations in these systems. However, neutron diffraction studies have yet to fully elucidate the magnetic structures of Cr$_{1+\delta}$Te$_2$ compounds. 
\par
\begin{figure}[ht]
\centering
\includegraphics[width=8cm,height=7cm]{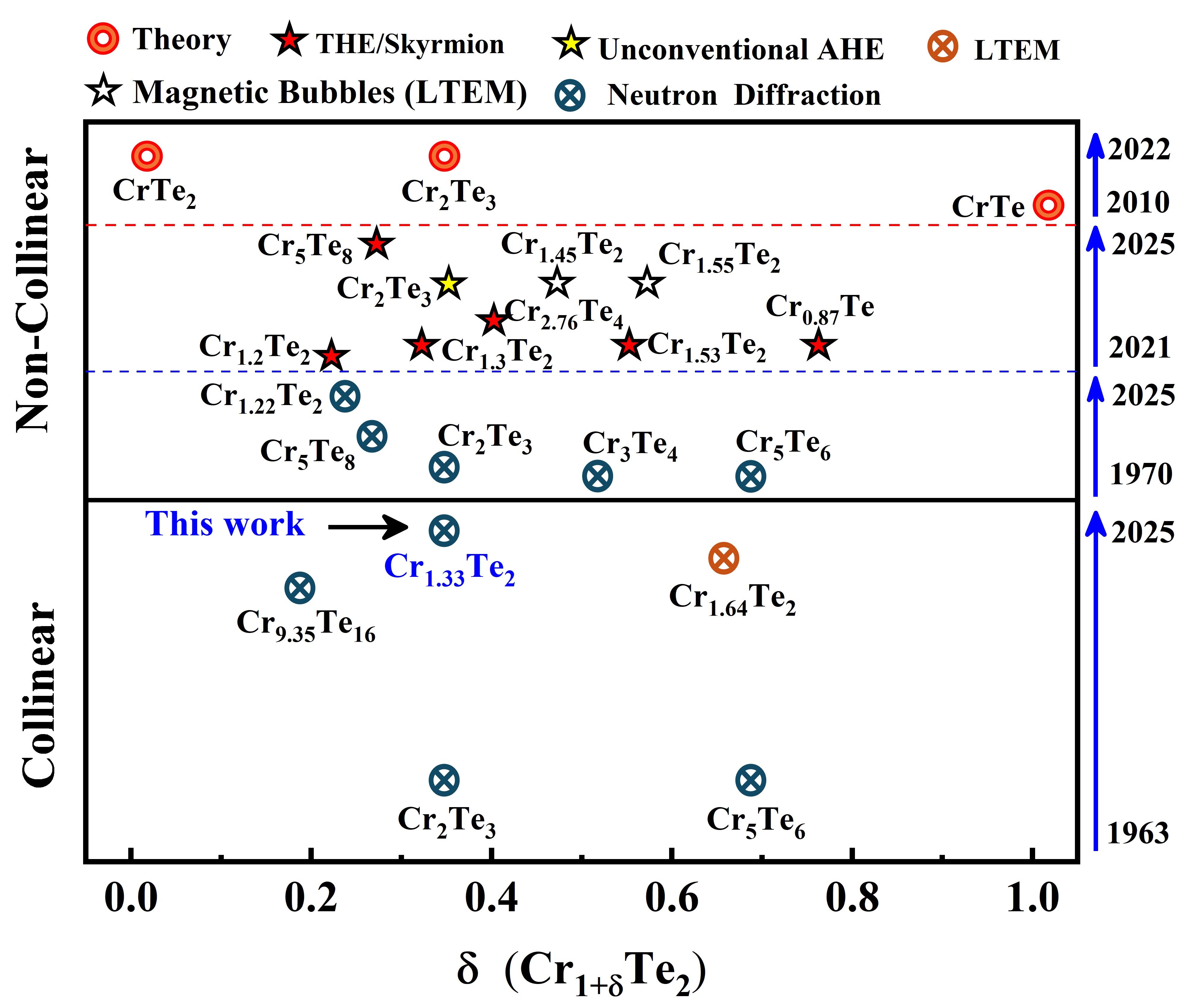}
\caption{Classification of magnetic structures in Cr-Te systems: collinear and non-collinear, based on theoretical and experimental study.}
\label{fig:Comparative_study}
\end{figure}

As shown in Fig.~\ref{fig:Comparative_study}, we categorize the magnetic structures of Cr-Te-based binary compounds into two groups: collinear and non-collinear. Non-collinear structures are identified through theoretical predictions or experimental evidence, such as {\sf THE}, skyrmion observations, neutron diffraction, or unconventional Anomalous Hall Effect (AHE)~\cite{Cr2Te3_Theory_GS, CrTe2_Theory, CrTe_Theory, Cr0.87Te_ACS_Nano, Cr1.53Te2_Adv, Cr5Te8_Nitesh, Cr1.2Te2_Nano_Letter, Cr2.76Te4, Cr1.61Te2_THE, Cr1.33Te2_Rana_Saha, CrTe2_AdvF, Cr2Te3_Unconvetional_AHE, Cr1.22Te2, Cr5Te8_nonCo, Cr3Te4_Cr5Te6_nonCo, Cr2Te3_nonCo}. In contrast, collinear structures are primarily determined by neutron diffraction and LTEM~\cite{CrTe2_AdvF, Cr12-xTe16, Cr2Te3_Cr5Te6_collinear}. In particular, Fig.~\ref{fig:Comparative_study} reveals no clear correlation between the composition parameter $\delta$ and the magnetic structure. To date, neutron diffraction studies~\cite{Cr12-xTe16, CrTe2_AdvF, Cr2Te3_Cr5Te6_collinear,Cr1.22Te2, Cr5Te8_nonCo, Cr3Te4_Cr5Te6_nonCo, Cr2Te3_nonCo} on Cr$_{1+\delta}$Te$_2$ have been limited, mostly conducted on powder samples, and the results for samples of identical composition often vary, as illustrated in Fig.~\ref{fig:Comparative_study}. Thus, single-crystal neutron diffraction studies of Cr$_{1+\delta}$Te$_2$ are crucial for resolving these inconsistencies and clarifying its magnetic configurations.

\par
Our neutron diffraction study on Cr$_{1.33}$Te$_2$, previously reported to show a N\'eel-type skyrmion and {\sf THE}~\cite{Cr1.33Te2_Rana_Saha}, unexpectedly revealed a collinear magnetic structure. EXAFS analysis revealed vacancies at Cr and Te sites, which, combined with DFT calculations, indicated that orbital degeneracies significantly influence the magnetocrystalline anisotropy energy (MAE)~\cite{PhysRevB.100.144413, PhysRevResearch.4.013237, PhysRevB.81.104426}. In a vacancy-free system, orbital degeneracy favors a non-collinear spin arrangement, but vacancies disrupt the Cr-Te bonding, lifting $d$-orbital degeneracy and stabilizing a collinear magnetic state.



\vskip 0.15cm
\noindent\textit{Single crystal neutron diffraction:} High quality single crystals of Cr$_{1+\delta}$Te$_2$ ($\delta \approx ~$0.33), used in this study, were synthesized by chemical vapor transport method (CVT)~\cite{PRM_Cr1.33Te2}. To accurately determine the magnetic ground state of Cr$_{1+\delta}$Te$_2$ and obtain a reliable magnetic propagation vector, temperature ($T$) dependent single-crystal neutron diffraction was performed on the SXD instrument~\cite{SXD} at the ISIS Facility, UK, using time-of-flight Laue technique to map the three-dimensional reciprocal space (see Appendix~\ref{Experimental techniques_A}). Fig.~\ref{Laue} (a)-(c) show the neutron diffraction patterns in the $(h~k~0)$ scattering plane collected at $T =$ 300 K, 140 K, and 4.5 K, respectively (180 and 90 K data given in Fig.~\ref{Laue_supple} of Appendix~\ref{Laue neutron diffraction pattern_C}). The diffraction pattern at $T =$ 300 K exhibits only the nuclear reflections permitted by the lattice symmetry of the \textit{P}$\overline{3}$m1 space group (No. 164). Upon cooling below the transition temperature ($T_C =$ 191 K), a distinct change in the intensity of several Bragg reflections is observed, arising from the onset of long-range magnetic order; selected reflections exhibiting intensity enhancement are marked by red circles Fig.~\ref{Laue}. All magnetic reflections sit on the nuclear peak positions, indicating a commensurate magnetic structure with the propagation vector $\mathbf{k}=$ (0, 0, 0). 

\begin{figure*}
\centering
\includegraphics[width=16cm, height=5.5cm]{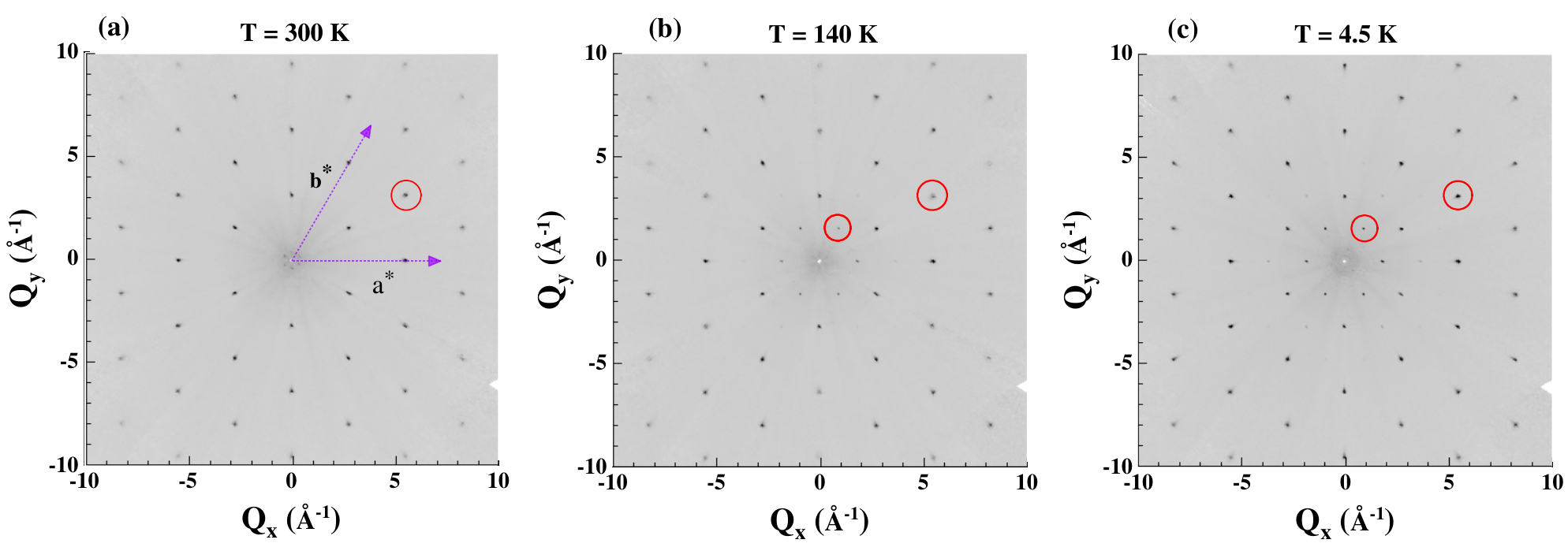}
\caption{(a), (b), and (c) show the $(h~k~0)$ layer of Cr$_{1+\delta}$Te$_2$ measured on SXD at $T = 300$, 140, and 4.5 K, respectively. Laue symmetry $\overline{3}m$ has been applied.}
\label{Laue}
\end{figure*}

\par

\begin{figure}[ht]
\centering
\includegraphics[width=8cm,height=7cm]{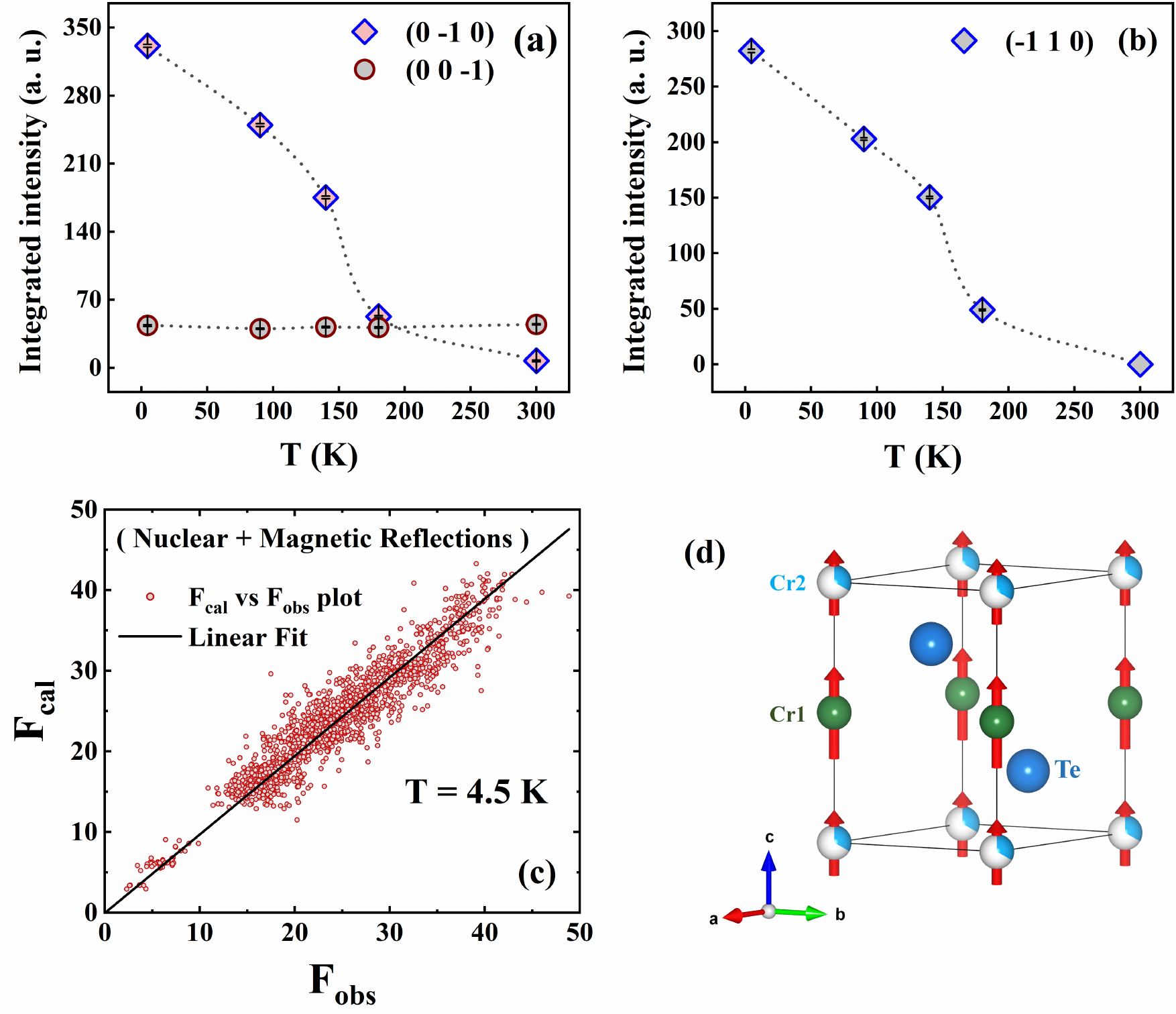}
\caption{(a), (b) Temperature evolution of integrated intensity of some selected Bragg peaks (0 -1 0), (0 0 -1) and (-1 1 0). (c) Calculated (F$_{cal}$) vs observed (F$_{obs}$) structure factors with linear fit at $T=$ 4.5 K (nuclear+magnetic reflections) respectively. (d) Magnetic structure of Cr$_{1+\delta}$Te$_2$ at $T=$ 4.5 K.}
\label{Intensity}
\end{figure}

To elucidate the magnetic moment orientation, we examined the $T$ dependence of the integrated intensities of the (0 $\overline{1}$ 0), (0 0 $\overline{1}$), and ($\overline{1}$ 1 0) Bragg peaks [Fig.~\ref{Intensity}]. The intensities of the (0 $\overline{1}$ 0) and ($\overline{1}$ 1 0) peaks increase upon cooling, following the $M$-$T$ curve [Fig.~\ref{Neutron D} (a) of Appendix~\ref{Temperature dependent magnetic data_F}], while the (0 0 $\overline{1}$) peak remains nearly $T$ independent. Since neutron scattering is sensitive only to the component of the magnetic moment perpendicular to the scattering vector, this behavior indicates that the magnetic moments are aligned along the $c$-axis. The result indicates the absence of appreciable spin canting upon cooling, contrasting with the canted magnetic structure reported for Cr$_{1.22}$Te$_2$~\cite{Cr1.22Te2}, but remains consistent with the collinear magnetic structure observed in Cr$_{12-x}$Te$_{16}$~\cite{Cr12-xTe16}.

\par
Based on the crystal symmetry and magnetic propagation vector $\mathbf{k}$, only four symmetry-allowed magnetic configurations are possible for Cr$_{1+\delta}$Te$_2$, as summarized in Table~\ref{table:MAXMAGN} of Appendix~\ref{Symmetry allowed magnetic structures_D}. Among these, only the magnetic subgroup \textit{P}$\overline{3}$m$'$1 (\#164.89) permits a finite magnetic moment along the $c$-axis, consistent with the experimental observation. Consequently, the parent space group \textit{P}$\overline{3}$m1 was used for nuclear refinement, and \textit{P}$\overline{3}$m$'$1 for the combined nuclear and magnetic refinement. The results of the refinement for $T=$4.5 K is presented as calculated (F$_{cal}$) versus observed (F$_{obs}$) structure factors with linear fits in Fig.~\ref{Intensity} (c). Further refinement details are presented in Fig.~\ref{Fobs_Fcal} of Appendix~\ref{Magnetic structure refinement_E}, and the corresponding refinement parameters and extracted magnetic moments are summarized in Table~\ref{table:Moment} of Appendix~\ref{Magnetic structure refinement_E}. We refine the moment of Cr1 and Cr2\footnote{In Cr$_{1+\delta}$Te$_2$, Cr has two inequivalent crystallographic sites, where Cr2 are the intercalated Cr with fractional occupancy residing between Cr1 and Te layers~\cite{PRM_Cr1.33Te2}.} separately at all the temperatures. At $T=$ 4.5 K, an ordered moment values of 2.13(5) $\mu_B$/Cr1 and 1.51(13) $\mu_B$/Cr2 are obtained along the $c-$axis and the schematic of the magnetic structure at $T=$ 4.5 K has been shown in Fig.~\ref{Intensity} (d). The $T$ dependence of the total ordered moment per unit cell, obtained from refinement, is shown in Fig. S3 (a)~\cite{SI}, alongside bulk magnetization data recorded under a 5 T applied field along the $c$-axis.   

\vskip 0.15cm
\noindent\textit{EXAFS:} In contrast to X-ray and neutron diffractions, which provide information on the long-range crystallographic structure, EXAFS offers direct insight into the local atomic environment. To obtain quantitative information on the local atomic structure of Cr$_{1+\delta}$Te$_2$, the Cr K-edge EXAFS spectra were analyzed over the $k$-range of 3 to 13.2 \AA$^{-1}$ ($k$ denotes the wavenumber of the photoelectrons). 

\begin{table}[htb]
    \caption{ The local structural parameters extracted from the EXAFS analysis at the Cr K-edge for $T$ = 300 K are summarized. Here, $N$ denotes the coordination number and $\sigma^2$ the Debye-Waller factor. The interatomic distances obtained from EXAFS fitting and from XRD refinement are listed as $R$ and $R_{\text{XRD}}$, respectively. Parameters that were fixed or constrained during the fitting are indicated by the hash sign (\#). The $R$-factor, which quantifies the degree of mismatch between the experimental data and the best-fit model, was found to be 0.0029.}
    \label{table:EXAFS_table}
    \centering
    \setlength{\tabcolsep}{3.8pt} 
    \renewcommand{\arraystretch}{1.8} 
    \begin{tabular}{c   c   c   c   c  c}
        \hline
        \hline
        Path & N$^{\#}$ & $\sigma^2$(\AA$^2$) & R(\AA) & R$_{XRD}$(\AA) &   \\
        \hline
        Cr1-Te & 4.5 & 0.0058$\pm$0.0003  & 2.720$\pm$0.003 & 2.732$\pm$0.001  &   \\
        Cr1-Cr2 & 0.5 &  0.0064$\pm$0.0055 & 2.937$\pm$0.04 & 3.028$\pm$0.000 & \\
        Cr1-Cr1 & 4.5 &  0.0221$\pm$0.0036 & 3.898$\pm$0.021 & 3.930$\pm$0.000 & \\
        
        \hline
    \end{tabular}

\end{table}

\par
\begin{figure}[t]
	\centering
	\includegraphics[width=8.7cm,height=6.0cm]{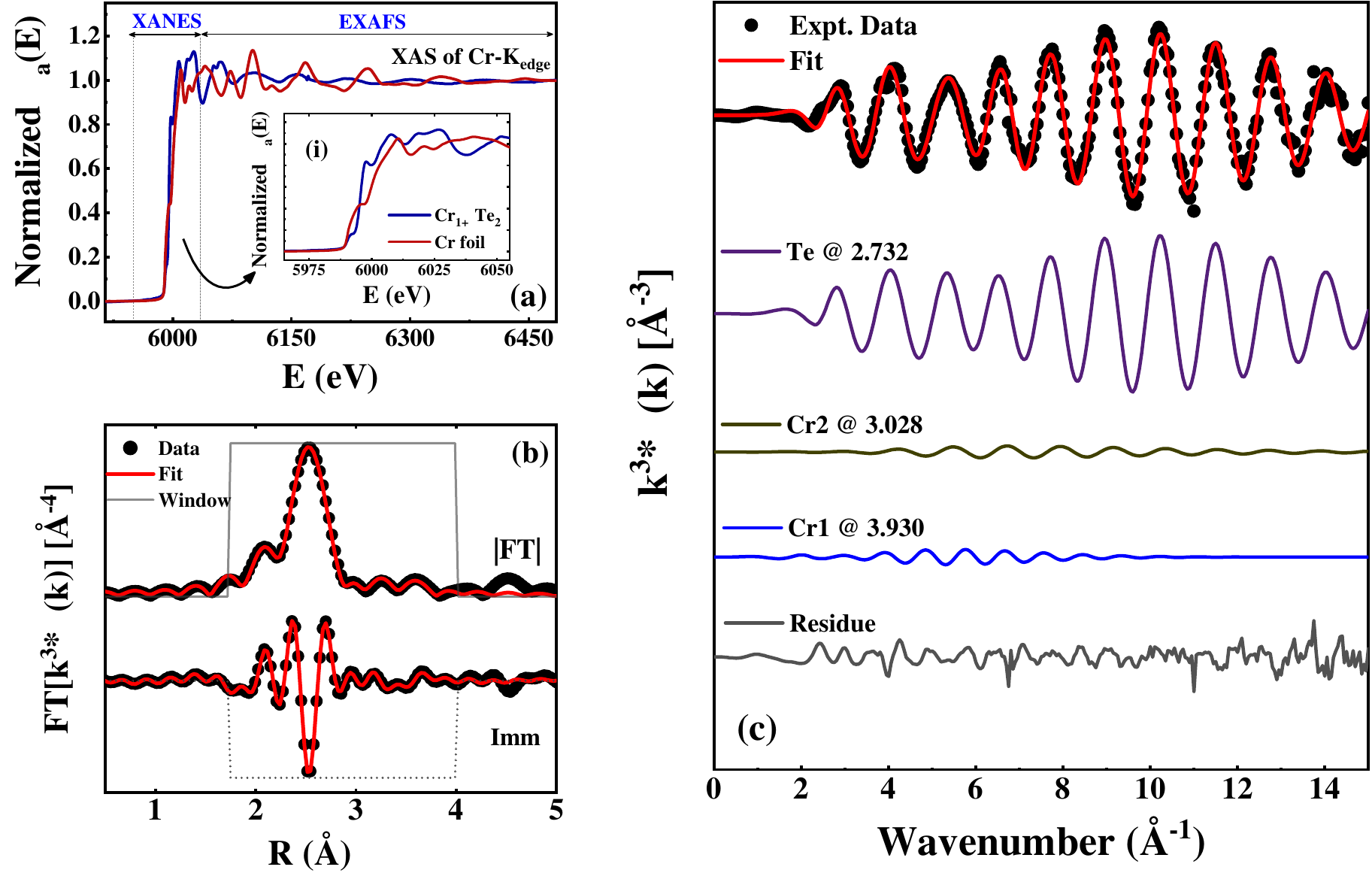}
	\caption{ (a) Normalized XAS spectra of Cr$_{1+\delta}$Te$_2$ and Cr foil measured at the Cr K-edge. The inset (i) shows the enlarge view of the XANES region. (b) Fourier transform of the experimental EXAFS data (black circles) along with the best-fit curve (solid red line). The magnitude ($|FT|$) and imaginary component ($I{\text{mm}}$) have been labeled and vertically shifted for clarity. (c) $k^3$-weighted Cr K-edge EXAFS spectra in $k$-space (black circles) together with the corresponding best-fit curve (solid red line). The contributions from individual single-scattering paths, labeled by the scattering atom and its distance from the absorber.}
	\label{EXAFS Ch4}
\end{figure}

Fig.~\ref{EXAFS Ch4}(a) shows the normalized XAS spectra [$\mu_a(E)$ vs. $E$\footnote{$E$ is the energy of the incident x-ray photon, and $\mu_a(E)$ is the x-ray absorption coefficient.}] of Cr$_{1.33}$Te$_2$, together with that of Cr foil, measured at the Cr K-edge and the inset shows an enlarged view of the XANES (X-ray Absorption Near Edge Structure) region. The rising line of the XANES spectra of Cr$_{1.33}$Te$_2$ is found to be shifted to higher energy compared to the data for the metallic Cr foil, indicating that Cr is in the higher oxidization state in Cr$_{1.33}$Te$_2$. Fig.~\ref{EXAFS Ch4}(b) presents the Fourier transform of the EXAFS oscillations, displaying both the imaginary component and the modulus ($|FT|$), along with the corresponding best-fit curve in $R$-space. The EXAFS data were fitted using a multi-shell refinement procedure up to 4 \AA\ to extract structural information around Cr, covering the first three coordination shells. The scattering paths used to generate the theoretical spectra were calculated from the crystallographic structure of Cr$_{1+\delta}$Te$_2$ obtained from diffraction studies. The effective scattering amplitudes and phase shifts of the atoms were computed using the FEFF6 code~\cite{FEFF6} within the Artemis software package. Since two crystallographically distinct Cr sites are present, namely, Cr1 (fully occupied) and Cr2 (partially occupied), only the single-scattering paths involving nearest-neighbour Cr1-Te, next-nearest-neighbour Cr1-Cr2, and next-next-nearest-neighbour Cr1-Cr1 were considered in the analysis. The results of the EXAFS fitting are summarized in Table~\ref{table:EXAFS_table}. The $k^3$-weighted best-fit spectra in $k$-space, together with the contributions from the individual single-scattering paths, are shown in Fig.~\ref{EXAFS Ch4}(c). In the figure, the scattering paths are labeled according to the scattering atom and its distance from the absorber. For example, Te@2.732 indicates that the scattering atom is Te and it is at at distance of 2.732~\AA~ from the absorbing atom (here Cr1).

\par
It is evident from Fig.~\ref{EXAFS Ch4}(b) and (c) that the EXAFS signal is predominantly governed by the scattering contribution from the first nearest neighbour, namely the Cr1-Te single-scattering path. A key outcome of the present EXAFS analysis is the identification of vacancies at both Te and Cr sites, as summarized in Table~\ref{table:EXAFS_table}. In the ideal crystal structure of Cr$_{1+\delta}$Te$_2$, each Cr1 atom is expected to be coordinated by six Te atoms in its first coordination shell; however, in the present case the coordination number is reduced to approximately 4.5 indicating significant Te vacancies. In the second coordination shell, the expected coordination number of about 0.66 for Cr2 atoms (arising from their fractional occupancy) is found to decrease to roughly 0.5. Similarly, vacancies are also evident in the third coordination shell.

\begin{figure*}[t]
	\centering
	\includegraphics[width=17.5cm, height=8cm]{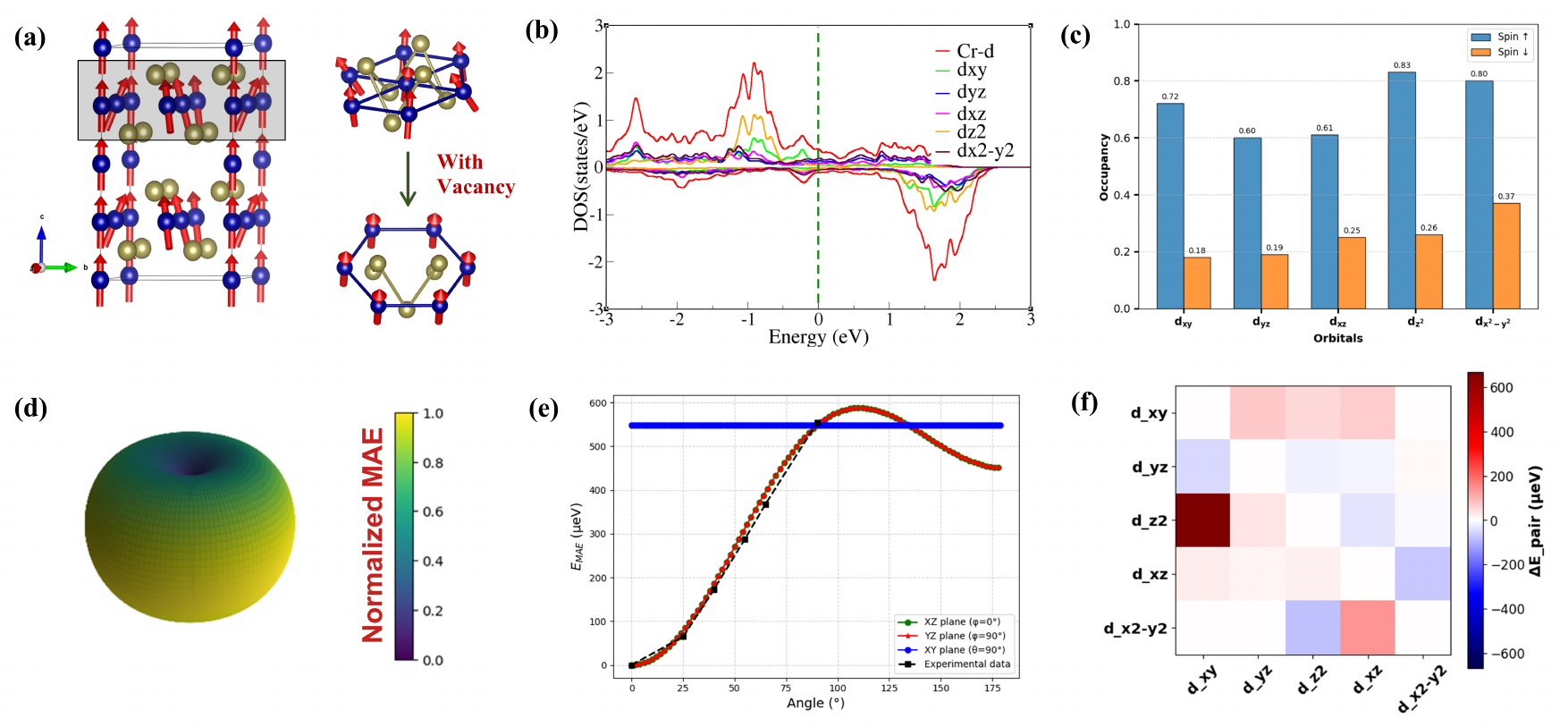}
	\caption{(a) Magnetic configurations of Cr$_{10}$Te$_{16}$ (Cr$_{1.25}$Te$_2$) and vacancy-containing Cr$_8$Te$_{12}$ (Cr$_{1.33}$Te$_2$). The upper-right panel shows the zoomed layer of the vacancy-free system, while the lower panel depicts the favorable configuration with vacancy. (b) Partial density of states (PDOS) of Cr $d$ orbitals for spin-up and spin-down channels. (c) Occupancy of different $d$ orbitals for spin-up and spin-down channels. (d) Normalized magnetocrystalline anisotropy energy (MAE) in spherical polar coordinate space. (e) MAE in different planes as a function of the polar angle, along with the experimental data presented with black symbols and dotted line. (f) Orbital-pair resolved contributions to the MAE in the presence of SOC.}
	\label{DFT1}
\end{figure*}
\vskip 0.15cm
\noindent\textit{Theoretical Study:} EXAFS analysis reveals the presence of vacancies in the experimentally investigated Cr$_{1+\delta}$Te$_2$ ($\delta \approx 0.33$) sample . Motivated by this, we systematically examined the magnetic ground state under two representative conditions: (i) a pristine lattice without vacancies (WoV) and (ii) a defected lattice incorporating vacancies (WV). Further structural information can be found in Appendix~\ref{Theoretical study_H}.
\par
For the WoV configuration, first-principle calculations~\cite{hafner2008ab,kresse1999ultrasoft,perdew} indicate that a non-collinear spin arrangement, tilted from the [0~0~1] direction at the 6$i$ site, is energetically preferred, yielding a total energy of $-150.7$ eV [Fig.~\ref{DFT1} (a), Fig.~\ref{Theory} of Appendix~\ref{Ground- and Excited-State Energy Landscapes of WoV and WV States_J}]. While the corresponding collinear configuration (spins strictly aligned along the [0~0~1] direction) lies marginally higher in energy by $0.1$ eV, establishing the non-collinear state as the true magnetic ground state of the WoV phase. 
\par
In striking contrast, the energetic hierarchy reverses in the WV system. The non-collinear configuration at the 6$i$ site no longer represents the ground state, exhibiting a total energy of $-106.5$ eV [Fig.~\ref{DFT1}(a) and  Fig.~\ref{Theory} of Appendix~\ref{Ground- and Excited-State Energy Landscapes of WoV and WV States_J}]. Instead, the collinear state becomes significantly stabilized, lying $2.2$ eV lower ($-108.7$ eV) than the non-collinear counterpart. This pronounced energy gap signifies a strong preference for collinear magnetism in the presence of vacancies. 
To gain microscopic insight into the orbital physics of the Cr atom at the 6$i$ site, we performed Wannierization~\cite{pizzi2020wannier90,MarzariPhysRevB.56.12847} to extract the onsite energies of the Cr $d$ orbitals. The resulting onsite energies $\epsilon_i$ (Table~\ref{table:onsite_energy} of Appendix~\ref{Onsite energy extraction from Wannier functions_K}) capture the effective orbital levels after incorporating the respective magnetic ground states. A close inspection of Table~\ref{table:onsite_energy} reveals a clear lifting of orbital degeneracy in the WV phase, which we attribute to symmetry reduction.
\par
In the WoV trigonal phase ($P\bar{3}m1$), Te atoms reside in octahedral coordination with Cr (Fig.~\ref{Octahedra} of Appendix~\ref{Octahedral geometry_L}). The threefold rotational symmetry along the $c$ axis reduces the cubic $O_h$ crystal field to $D_{3d}$, lifting the degeneracy of the $t_{2g}$ ($d_{xy}$, $d_{yz}$, $d_{xz}$) manifold while leaving the $e_g$ doublet ($d_{z^{2}}$, $d_{x^{2}-y^{2}}$) nearly unaffected. In contrast, for the WV case, induced by Cr-Te bond rupture, the local symmetry constraints are completely removed. Consequently, the five $3d$ orbitals evolve into fully non-degenerate states, exhibiting hybridized $d_{z^{2}}$, $d_{x^{2}-y^{2}}$, $d_{xy}$, $d_{xz}$, and $d_{yz}$ characters. To unravel how these orbitals interact with spin and determine its orientation, we next analyze the magnetocrystalline anisotropy of the system. The MAE was evaluated within the second-order perturbation theory using the spin-orbit coupling (SOC) Hamiltonian~\cite{Judd1963,Cooper1979,Bruno1989}, as given in Eq.~\ref{SOC}, where $\mathbf{L}$ and $\mathbf{S}$ denote the orbital and spin angular momenta of Cr.

\begin{equation}
H_{\mathrm{SOC}} = \xi  \mathbf{L}\cdot \mathbf{S}
= \xi \left( L_x S_x + L_y S_y + L_z S_z \right),
\label{SOC}
\end{equation}
where $\xi$ denotes the SOC strength (0.0130~eV)~\cite{song2022superexchange,koseki2019spin}, $L_{x,y,z}$ are orbital angular momentum matrices in the $d$-orbital basis ($l=2$), and $S_{x,y,z}$ are spin-$\tfrac{1}{2}$ operators.

Within this framework, the second-order SOC correction of the total energy for a spin quantization axis $\hat{n}$ is expressed as
\begin{equation}
E^{(2)}(\hat{n}) = - \sum_{i \in \mathrm{occ}} \sum_{j \in \mathrm{unocc}}
\frac{ \big| \langle \psi_j | H_{\mathrm{SOC}}(\hat{n}) | \psi_i \rangle \big|^2 }{\varepsilon_j - \varepsilon_i},
\label{2nd SOC}
\end{equation}
Detailed derivations are provided in Appendix~\ref{Perturbative Spin-Orbit Coupling Formalism for MAE_M}. The spin moment is governed by orbital occupancy, which scales with the numerator of Eq.~\ref{2nd SOC}. Orbital-resolved projected density of states (PDOS) calculations reveal that the redistribution of $d$-orbital occupations arises from the interplay between Hund’s exchange and the on-site Coulomb repulsion (Hubbard $U$) [see Appendix~\ref{Computational details_I}]. The PDOS [Fig.~\ref{DFT1} (b)] shows clear exchange splitting between spin-up and spin-down $d$ orbitals, characteristic of a ferromagnetic ground state. Integration of the PDOS yields orbital occupancies [Fig.~\ref{DFT1} (c)], displaying distinct filling asymmetries between the spin channels. Subsequently, SOC energies were computed for magnetization along the [0~0~1] ($z$) and [0.25~ –0.25~ 1.00] ($\hat{m}$) directions using Eq.~\ref{MAE}.

The MAE is obtained as
\begin{equation}
\mathrm{MAE} =  E^{(2)}(\hat{m}) -E^{(2)}(z),
\label{MAE}
\end{equation}
where a positive (negative) value corresponds to an easy axis along the $z$ ($\hat{m}$) directions. The resulting magnetic anisotropy energy is $\mathrm{MAE} \approx 550~\mu\text{eV}$ [evaluated at $\theta = 90^\circ$; see Fig.~\ref{DFT1}(e)], indicating an easy axis along $z$; however, in the WoV system, the sign reverses, signifying a reorientation of the easy axis toward $\hat{m}$ [a detailed discussion is provided in Appendix~\ref{Perturbative Spin-Orbit Coupling Formalism for MAE_M},~\ref{Hamiltonian Matrix Representation Before and After Spin–Orbit Coupling_N},~\ref{MAE Characteristics of the WoV System_O}]. 

To construct the complete anisotropy landscape, we evaluated the SOC energies for magnetization orientations parameterized in spherical coordinates,
[
$\hat{S}(\Theta,\Phi) = \big(\sin\Theta\cos\Phi,~\sin\Theta\sin\Phi,~\cos\Theta\big)$
]
with $\Theta \in [0,\pi]$ and $\Phi \in [0,2\pi)$. The corresponding SOC energy is given by
[
$E^{(2)}(\Theta,\Phi) = E^{(2)}\big(\hat{S}(\Theta,\Phi)\big)$.
]
For visualization, the energies were normalized between 0 and 1. Fig.~\ref{DFT1} (d) presents the normalized 3D MAE surface, where the color gradient approaches blue near the $z$-axis, indicating a diminishing MAE magnitude.
\par
To further elucidate the anisotropy behavior, we examined 2D cross-sections of the 3D MAE surface along the principal crystallographic planes ($x$–$z$, $y$–$z$, and $x$–$y$). As shown in Fig.~\ref{DFT1}(e), the MAE remains nearly constant at $\sim$550~$\mu$eV across the $x$–$y$ plane, while the energy minimum aligns with the $z$-axis. The measured MAE as a function of angle $\theta$, extracted from the $M$-$H$ curves at $T = 2.5$~K, is in excellent agreement with the theoretical MAE curve for the $x$-$z$ plane [Fig.~\ref{DFT1}(e)]. At $\theta = 90^\circ$, our measurements yield a MAE of 554~$\mu$eV, corresponding to an effective anisotropy constant of $\text{K}_{\mathrm{eff}} = 14.6~\text{J\,m}^{-3}$. This value of $\text{K}_{\mathrm{eff}}$ is consistent with those reported for bulk single crystals of Cr$_{1+\delta}$Te$_2$~\cite{Cr1.53Te2_Adv} (see Appendix~\ref{Perpendicular magnetic anisotropy_G}). The microscopic origin of this anisotropy can be rationalized within the second-order perturbative framework, where SOC mediates virtual excitations between crystal-field–split orbitals. The pairwise anisotropy~\cite{Bruno1989} contribution is expressed as
\begin{equation}
E_{ij} =
\frac{f_i (1 - f_j)}{\varepsilon_i - \varepsilon_j}
\left| \langle i | H_{\text{SOC}} | j \rangle \right|^2,
\label{eq:orbital_MAE}
\end{equation}
where $f_i$ and $\varepsilon_i$ are the occupation and crystal-field energy of orbital $i$, and
$\langle i | H_{\text{SOC}} | j \rangle$ represents the SOC matrix element between orbitals $i$ and $j$.
This formulation captures how SOC-driven inter-orbital hybridization governs the anisotropy landscape, as illustrated in Fig.~\ref{DFT1} (f).

\begin{table}[h!]
\centering
\caption{Orbital-pair contributions (axis = [0, 0, 1])}
\label{tab:orbital_MAE}
\begin{tabular}{cccccc}
\hline
\textbf{orb\_i} & \textbf{orb\_j} & $\mathbf{\Delta\varepsilon}$ (eV) & $\mathbf{f_i(1-f_j)}$ & $\mathbf{|H_n|}$ (meV) & $\mathbf{E_{\text{pair}}}$ ($\mu$eV) \\
\hline
$d_{xy}$ & $d_{z^2}$     &  0.050 & 0.205 & 1.806 &   31.956 \\
$d_{xy}$ & $d_{xz}$      & -0.150 & 0.257 & 1.475 &  -65.164 \\
$d_{yz}$ & $d_{x^2-y^2}$ & -0.550 & 0.164 & 1.475 &   -7.576 \\
$d_{z^2}$ & $d_{xy}$     & -0.050 & 0.300 & 1.806 & -710.751 \\
$d_{z^2}$ & $d_{x^2-y^2}$& -0.350 & 0.226 & 1.806 &   -7.758 \\
$d_{xz}$ & $d_{xy}$      &  0.150 & 0.237 & 1.475 &   39.637 \\
$d_{x^2-y^2}$ & $d_{yz}$ &  0.550 & 0.354 & 1.475 &   18.968 \\
$d_{x^2-y^2}$ & $d_{z^2}$&  0.350 & 0.266 & 1.806 &   88.318 \\
\hline
\end{tabular}
\end{table}

The orbital-pair decomposition (Table~\ref{tab:orbital_MAE}) reveals that the dominant contribution arises from the
$ d_{xy} \leftrightarrow d_{z^2} $ coupling, characterized by a small energy separation
$\Delta \varepsilon \approx $~-0.05~\text{eV} and a strong SOC matrix element $|H_n| \approx$ 1.8~\text{meV}. This strong hybridization between the in-plane ($ d_{xy}$) and out-of-plane ($d_{z^2}$) orbitals generates the most substantial negative anisotropy term, favoring magnetization along the $z$-axis. In contrast, couplings involving the $d_{xz}$, $d_{yz}$ and $d_{x^2-y^2}$ orbitals yield weaker contributions due to larger crystal-field splittings and reduced SOC-induced mixing. The absence of diagonal terms confirms that anisotropy arises purely from inter-orbital interactions [white region along diagonal in Fig.~\ref{DFT1} (f)]. Collectively, these results demonstrate that the pronounced $d_{xy} \leftrightarrow d_{z^2}$ hybridization underlies the strong out-of-plane anisotropy, stabilizing magnetization along the [0~0~1] direction. This orbital-selective SOC mechanism offers a microscopic origin for the easy-axis behavior and underscores the pivotal role of specific $3d$ orbital couplings in governing the magnetic anisotropy of Cr$_{1+\delta}$Te$_2$.
\vskip 0.15cm
\noindent\textit{Summary and Conclusions:} Our single-crystal neutron diffraction study reveals that Cr$_{1.33}$Te$_2$ stabilizes in a collinear ferromagnetic structure, despite earlier reports observing N\'eel-type skyrmions and THE for this composition. Despite a seemingly well-ordered average structure, Cr K-edge EXAFS reveals the presence of both Te and Cr vacancies in the local environment. Cr$_{1.33}$Te$_2$ exhibits strong perpendicular magnetic anisotropy with easy $c$-axis, and together with theoretical calculations we found that vacancies play a decisive role in tuning the magnetic anisotropy and stabilizing the ground state magnetic structure from non-collinear to collinear. In conclusion, these findings establish a broadly applicable paradigm for engineering magnetocrystalline anisotropy and can help resolve the long-standing ambiguity in the magnetic structure of Cr-Te-based vdW systems. By exploiting crystal symmetries-such as the octahedral coordination presented here-to generate quasi-degenerate orbital manifolds, one can precisely tailor the MAE through controlled lifting of orbital degeneracy. This approach offers a unified framework for designing anisotropic magnetic behavior via symmetry-guided orbital engineering.

\vskip 0.25cm

\noindent\textit{Acknowledgments:} P.C. gratefully acknowledges the DST-INSPIRE program (Grant No. DST/INSPIRE Fellowship/2019/IF190532) for research assistance. We are thankful for the India-RAL collaborative project (Project No. SR/NM/Z-07/2015) for the neutron scattering studies at the ISIS facility, RAL, UK (Grant No. RB2410336~\cite{Cr1deltaTe2_neutron_2025}). We also acknowledge Elettra Sincrotrone Trieste for providing access to its synchrotron radiation facilities with XAFS beamline. JS and MK acknowledge National Supercomputing Mission (NSM) for providing computing resources of ‘PARAM RUDRA’ at S.N. Bose National Centre for Basic Sciences, which is implemented by C-DAC and supported by the Ministry of Electronics and Information Technology (MeitY) and the Department of Science and Technology (DST), Government of India.

\newpage
\appendix
\section{Experimental techniques}
\label{Experimental techniques_A}
\vspace{0.2 cm}

The detailed method for the growth of Cr$_{1+\delta}$Te$_2$ ($\delta \approx ~$0.33) single crystals  was discussed in our previous paper~\cite{PRM_Cr1.33Te2}. To probe the magnetic structure, a single crystal neutron diffraction experiment was carried out at the Single Crystal Diffractometer (SXD) beamline installed at the ISIS spallation neutron source, where the time-of-flight Laue technique is used to access large three dimensional (3D) volume of reciprocal space in a single measurement~\cite{SXD}. For the experiment a single crystal of approximate dimension $10\times 2 \times 0.16$ mm$^3$ (mass $\approx$ 40 mg) was attached to an aluminum (Al) pin using thin strips of adhesive Al tape. The sample was mounted on a closed-cycle refrigerator inside the evacuated sample tank and data were collected at temperatures $T=$ 4.5, 90, 140, 180 and 300 K for several fixed crystal orientations. Data reduction was performed using SXD2001 software~\cite{SXD2001}. Nuclear and magnetic refinements were performed using JANA2020 software~\cite{Jana2020}. Magnetic measurements were carried out using the MPMS3 (SQUID magnetometer) from Quantum Design. Room-temperature X-ray absorption spectroscopy (XAS) at the Cr K-edge was carried out at the XAFS beamline of Elettra Sincrotrone Trieste, Italy~\cite{Elettra}. The incident X-ray energy was selected using a Si(111) double-crystal monochromator. The EXAFS region of the acquired XAS data was processed and analyzed using the open-source DEMETER software package, which includes ATHENA and ARTEMIS~\cite{EXAFS1,EXAFS2}.

\vspace{0.2 cm}
\section{Sample characterization} 
\label{Sample characterization_B}

\vspace{0.2 cm}
Cr$_{1.33}$Te$_2$ has been characterized through single crystal and powder x-ray diffraction, transmission electron microscopy (TEM), magnetization measurements (see~\cite{PRM_Cr1.33Te2}). The powder and single-crystal XRD analysis confirmed that Cr$_{1+\delta}$Te$_2$ crystallizes in a trigonal structure belonging to the space group \textit{P}$\overline{3}$m1 (No.~164) with no structural transition at low temperatures. The unit cell of Cr$_{1+\delta}$Te$_2$ contains two Cr sites, \emph{i.e.}, Cr1 and Cr2. The Cr2 site has fractional occupancy and is intercalated in the van der Waals gap between two Cr1Te$_2$ layers. Magnetic measurements with the applied field ($H$) oriented along the $c$-axis and within the $ab$-plane reveal strong magnetic anisotropy, with the $c$-axis serving as the easy axis of magnetization. The temperature ($T$) dependence of magnetization ($M$) reveals a paramagnetic (PM) to ferromagnetic (FM) transition at approximately 191 K for both orientations.

\vspace{0.2 cm}
\section{Laue neutron diffraction pattern} 
\label{Laue neutron diffraction pattern_C}

\begin{figure}[h]
\centering
\includegraphics[width = 12 cm]{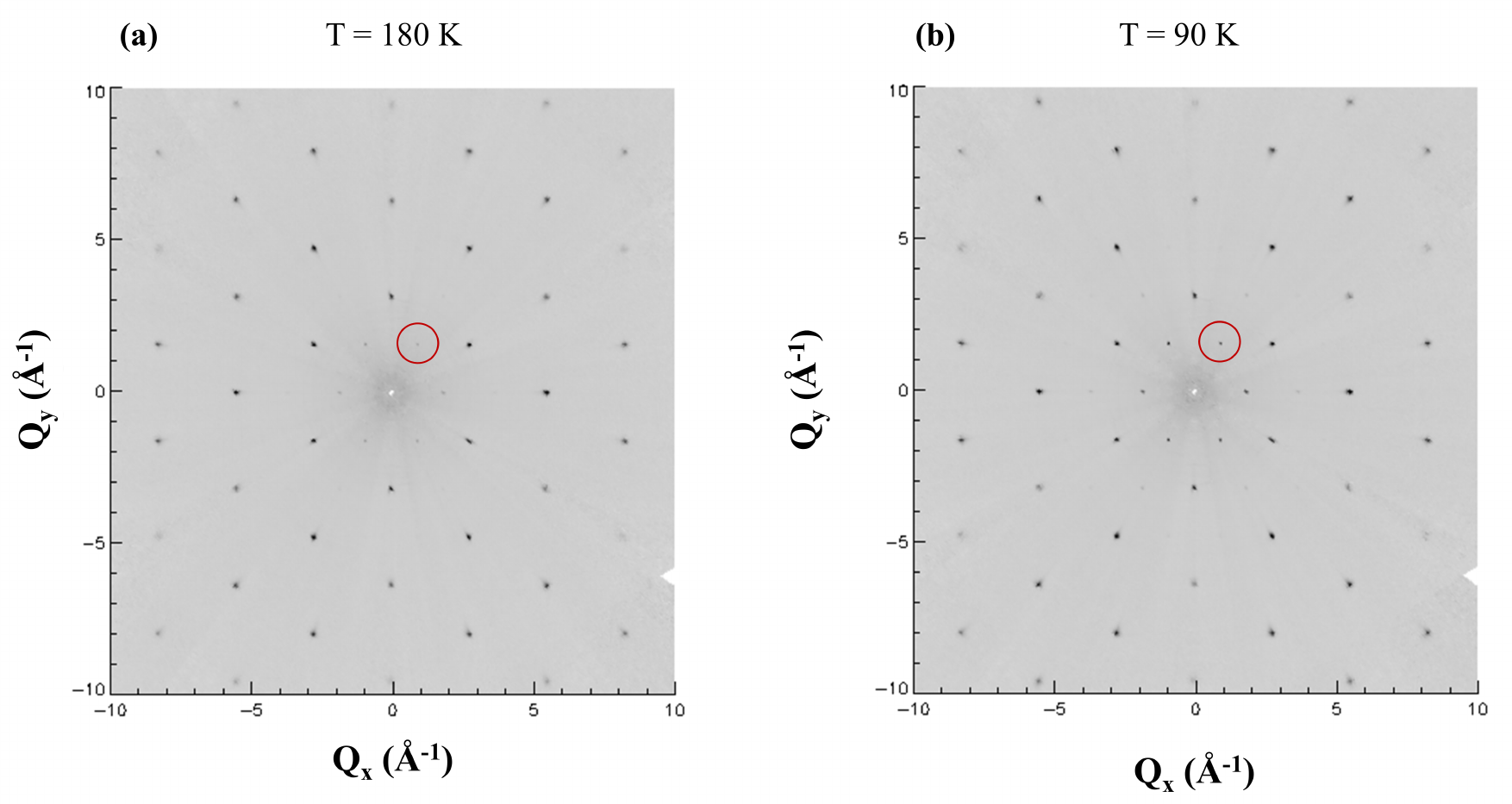}
\caption{\sffamily (a), and (b) show the $(h~k~0)$ layer of Cr$_{1+\delta}$Te$_2$ measured on SXD at $T =$ 180, and 90 K, respectively. Laue symmetry $\overline{3}m$ has been applied.}
\label{Laue_supple}
\end{figure}

\vspace{0.2 cm}
Fig.~\ref{Laue_supple}(a), and (b) show the neutron diffraction patterns in the $(h~k~0)$ scattering plane at $T =$ 180 K, and 90 K, respectively. The peaks marked with red circles show clear change in intensity with temperature ($T$).

\section{Symmetry allowed magnetic structures}
\label{Symmetry allowed magnetic structures_D}
\vspace{0.2 cm}
\par
To determine the possible magnetic structures, a symmetry analysis was carried out using the MAXMAGN program available on the Bilbao Crystallographic Server~\cite{Bilbao,MAXMGN}. For this analysis, the crystallographic information file (CIF) of Cr$_{1+\delta}$Te$_2$ (as provided in Ref.~\cite{PRM_Cr1.33Te2}) was used, along with the experimentally observed magnetic propagation vector $\mathbf{k} = (0,~0,~0)$. The symmetry-allowed magnetic configurations corresponding to this setup are listed in Table~\ref{table:MAXMAGN}.

\begin{table}[h]
\centering
\caption{\sffamily Symmetry-allowed magnetic moments for Cr sites in different magnetic space groups. Moment components are given as $(M_x, M_y, M_z)$.}
\label{table:MAXMAGN}
\begin{tabular}{|c|c|c|c|c|}
\hline
   & \multicolumn{4}{|c|}{Allowed magnetic space groups} \\ 
\hline
 \shortstack{Magnetic \\ atoms} & \textit{P}$\overline{3}$m$'$1 & \textit{P}$\overline{3}'$m$'$1 & \textit{P}$\overline{3}'$m1 & \textit{P}$\overline{3}$m1 \\ 
 & (\#164.89) & (\#164.88) & (\#164.87) & (\#164.85) \\ 
\hline
Cr1 & (0, 0, $M_z$) & (0, 0, 0) & (0, 0, 0) & (0, 0, 0) \\
Cr2 & (0, 0, $M_z$) & (0, 0, 0) & (0, 0, 0) & (0, 0, 0) \\
\hline
\end{tabular}
\end{table}

\noindent From Table~\ref{table:MAXMAGN}, it is evident that only the magnetic space group \textit{P}$\overline{3}$m$'$1 permits a finite magnetic moment along the $z$ axis, whereas the others yield zero moment. Hence, \textit{P}$\overline{3}$m$'$1 was employed for the magnetic structure refinement.

\vspace{0.2 cm}
\section{Magnetic structure refinement}
\label{Magnetic structure refinement_E}
\vspace{0.2 cm}

\begin{figure}[h]
\centering
\includegraphics[width = 16 cm]{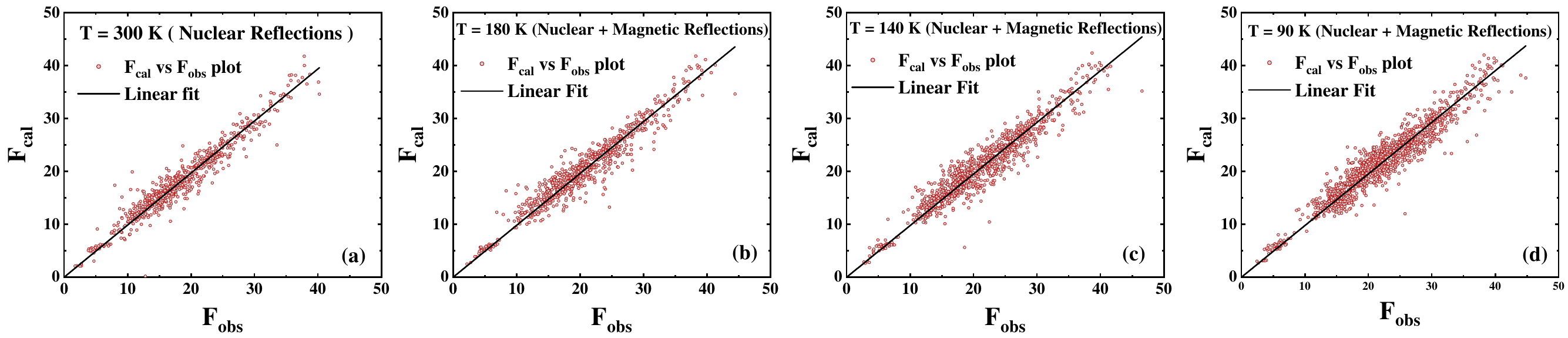}
\caption{\sffamily (a)-(d) Calculated (F$_{cal}$) vs observed (F$_{obs}$) structure factors with linear fit at $T=$ 300 K, 180 K, 140 K, 90 K, respectively.}
\label{Fobs_Fcal}
\end{figure}

At $T=$ 300 K, the nuclear structure refinement has been done using the \textit{P}$\overline{3}$m1 space group and it is shown in Fig.~\ref{Fobs_Fcal} (a). Below $T_C$, the nuclear+magnetic structure refinement has been done using the space group \textit{P}$\overline{3}$m$'$1 and the corresponding plots are shown in Fig.~\ref{Fobs_Fcal} (b)-(d) for $T=$ 180K, 140K, 90K. The ordered magnetic moments of the unit cell and individual Cr sublattices at different temperatures are listed in Table~\ref{table:Moment}.

\begin{table}[t]
\centering
\caption{\sffamily Lattice parameters, refinement parameters and obtained magnetic moments (only $M_z$ values)}
\label{table:Moment}
\resizebox{\columnwidth}{!}{%
\begin{tabular}{|c|c|c|c|c|}
\hline
 Temperature (K) & \shortstack{Lattice \\ parameters (in \AA)} & R-Factors & \shortstack{Individual Cr moments \\ ($M_z$ in $\mu_B$)} & \shortstack{Total moment in\\   Unit Cell ($M_z$ in $\mu_B$)} \\ 
\hline
4.5 & \shortstack{a=b=3.946 \\ c=5.998} & \shortstack{\emph{G.O.F}=2.71 \\ R=7.62\%, wR2=14.36\%} & \shortstack{Cr1=2.13(5) \\ Cr2=1.51(13)} & 2.629(19) \\
\hline
90 & \shortstack{a=b=3.944 \\ c=6.006} & \shortstack{\emph{G.O.F}=2.86 \\ R=7.36\%, wR2=14.32\%} & \shortstack{Cr1=2.00(5) \\ Cr2=1.05(14)} & 2.352(19) \\
\hline
140 & \shortstack{a=b=3.941 \\ c=6.028} & \shortstack{\emph{G.O.F}=2.64 \\R=7.54\%, wR2=12.58\%} & \shortstack{Cr1=1.63(5) \\ Cr2=1.17(15)} & 2.02(2) \\
\hline
180 & \shortstack{a=b=3.938 \\ c=6.039} & \shortstack{\emph{G.O.F}=2.76 \\R=7.11\%, wR2=12.27\%} & \shortstack{Cr1=0.57(8) \\ Cr2=1.7(2)} & 1.14(3) \\
\hline
300 & \shortstack{a=b=3.935 \\ c=6.070} & \shortstack{\emph{G.O.F}=3.00 \\R=7.00\%, wR2=13.03\%} & - & - \\
\hline
\end{tabular}
}
\end{table}

\section{Temperature dependent magnetic data}
\label{Temperature dependent magnetic data_F}

\begin{figure}[h]
	\centering
	\includegraphics[width = 13 cm]{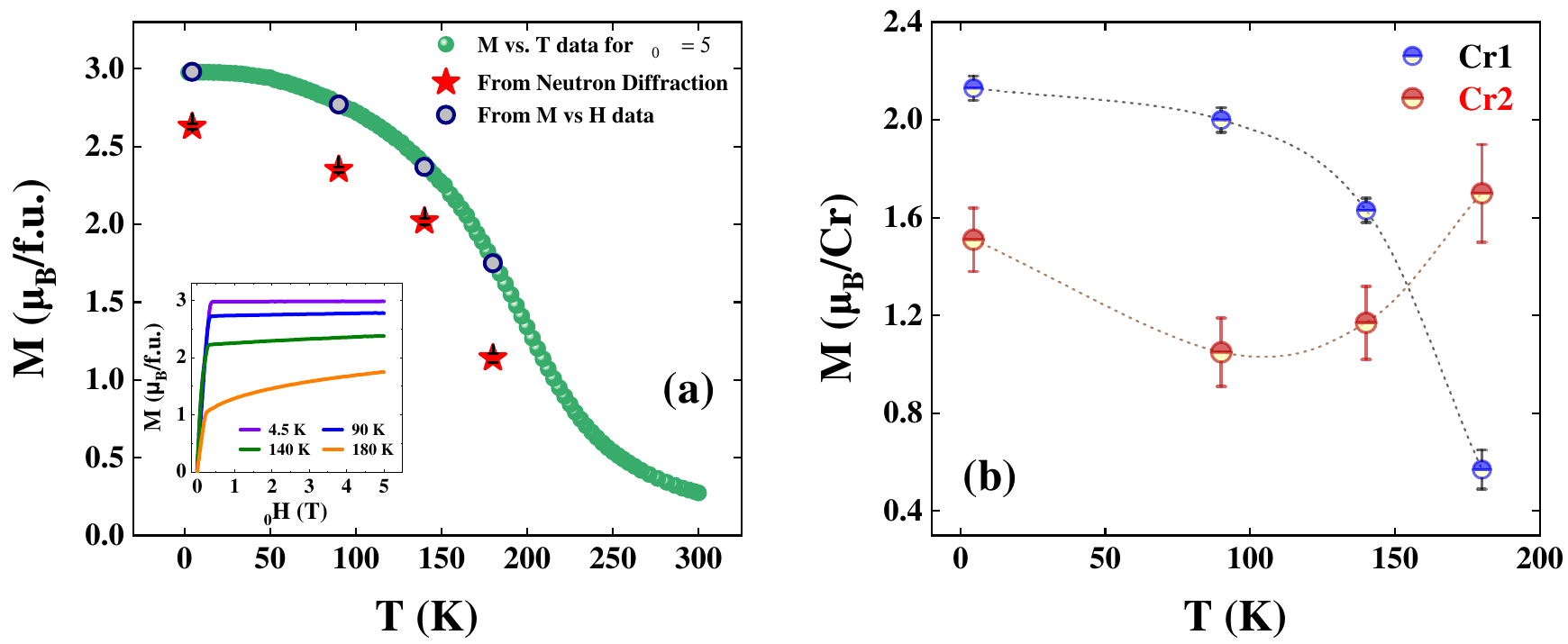}
	\caption{\sffamily (a) Main panel shows temperature variation of $M$ as obtained from single crystal neutron diffraction (total moment) and bulk magnetization measurements. $M$ vs $T$ data taken at 5 T and the inset shows the $M$ vs $H$ curve at $T=$ 4.5, 90, 140, 180 K (only 5 T values from it has been used in the main panel). (b) Temperature variation of ordered magnetic moment of Cr1 and Cr2 as obtained from neutron diffraction.}
	\label{Neutron D}
\end{figure}

\vspace{0.3 cm}
The $T$ dependence of the total ordered moment per unit cell, obtained from refinement, is shown in Fig.~\ref{Neutron D} (a), alongside bulk magnetization data recorded under a 5 T applied field along the $c$-axis. Since a single unit cell contains one formula unit (f.u.) of Cr$_{1+\delta}$Te$_2$, the ordered moment per unit cell is equivalent to that per formula unit. As shown in Fig.~\ref{Neutron D} (a), the total ordered moment per unit cell exhibits a temperature dependence that closely resembles that of the bulk magnetization. The inset of Fig.~\ref{Neutron D} (a) shows the $M$ vs. $H$ data recorded at $T$ = 4.5, 90, 140, 180 K. In Fig.~\ref{Neutron D} (b) the $T$ variation of ordered moment of Cr1 and Cr2 has been shown. Interestingly, one can see that moment of Cr1 site increases monotonically with decrease in $T$, while the moment of intercalated Cr2 site initially decreases from 180 K and then started increasing below 90 K.

\vspace{0.5 cm}
\section{Perpendicular magnetic anisotropy}
\label{Perpendicular magnetic anisotropy_G}
\begin{figure}[h]
	\centering
	\includegraphics[width = 15 cm]{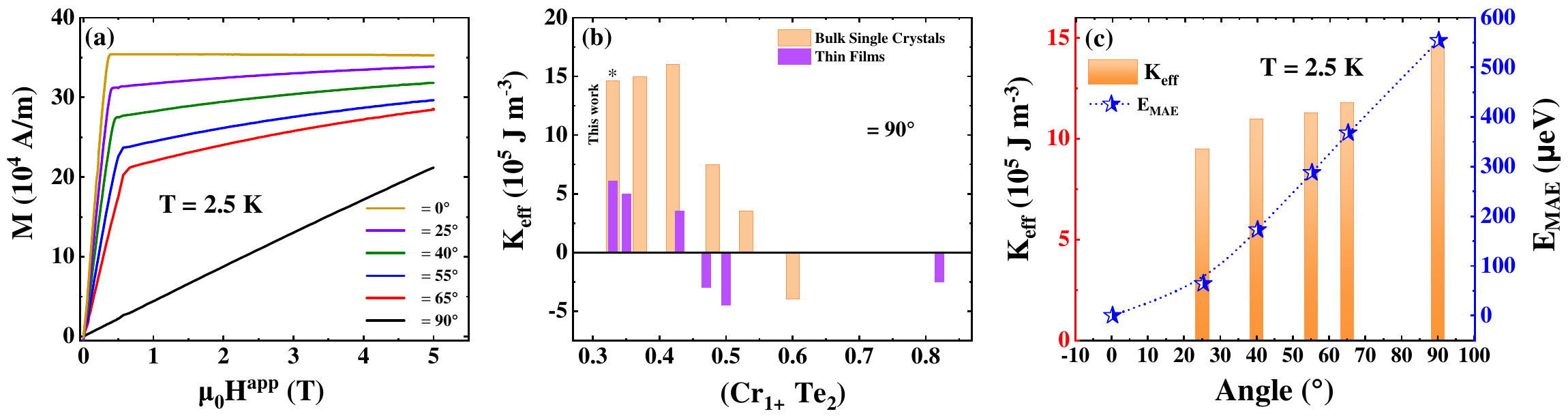}
	\caption{\sffamily (a) Field dependent magnetization curves measured at different $\theta$ values with respect to the $c$-axis at 2.5 K. (b) Variation of $\text{K}_{\text{eff}}$ with intercalated concentration $\delta$. The first column with star mark represent our calculated data and other data are taken from~\cite{Cr1.53Te2_Adv, Cr1+deltaTe2_PRM}. (c) Left panel shows $\theta$ variation of $\text{K}_{\text{eff}}$ and right panel shows $\theta$ variation magnetic anisotropy energy E$_{\text{MAE}}$.}
	\label{Anisotropy}
\end{figure}

\vspace{0.3 cm}
Fig.~\ref{Anisotropy} (a) shows the field-dependent magnetization curves, measured at different angles ($\theta$=0$^\circ$, 25$^\circ$, 40$^\circ$, 55$^\circ$, 65$^\circ$, 90$^\circ$) with respect to the $c$-axis at $T$=2.5 K. From the figure it is evident that the sample exhibits strong perpendicular magnetic anisotropy (PMA) with $c$-axis being the easy axis of magnetization, which has also been discussed in our previous paper~\cite{PRM_Cr1.33Te2}. When measuring or calculating magnetic anisotropy, the demagnetization factor (shape anisotropy) correction to the applied field ($\mu_0 H^{\mathrm{app}}$) should be considered to accurately determine the intrinsic anisotropy ($\emph{e.g.,}$ magneto-crystalline anisotropy). After correction, the effective magnetic field ($\mu_0 H^{\mathrm{eff}}$) inside the sample becomes:

\begin{equation}
    \mu_0H^{\mathrm{eff}}=\mu_0H^{\mathrm{app}}-\mu_0DM
\end{equation}

\noindent where $D$ is the demagnetization factor which is estimated to be $D_{ab}$=0.097 ($\mu_0 H^{\mathrm{app}}\parallel ab$, $\theta$=90$^\circ$) and $D_{c}$=0.8 ($\mu_0 H^{\mathrm{app}}\parallel c$, $\theta$=0$^\circ$). For $\theta$=25$^\circ$, 40$^\circ$, 55$^\circ$, 65$^\circ$ the D values are 0.6744, 0.5095, 0.3283, 0.2226 respectively. Now, if we have two $M$ vs. $H$ curves, one measured along the easy axis ($\theta=$0$^\circ$) and the other measured along different $\theta$ values, then the general anisotropy energy density difference between two directions can be expressed as: 

\begin{equation}
    \Delta E=\text{K}_\mathrm{eff}[\sin^2\theta-\sin^20^\circ]=\text{K}_\mathrm{eff}\sin^2\theta
\end{equation}

Experimentally, this energy difference equals the area between the two $M$ vs. $H$ curves:

\begin{equation}
   \Delta E=\mu_0\int_0^{M_s}[H^{\mathrm{eff}}(\theta;M)-H^{\mathrm{eff}}(0^\circ;M)]dM 
\end{equation}

Thus the effective magnetic anisotropy $\text{K}_\mathrm{eff}$ can be calculated using the following relation~\cite{Anisotropy_F1,Anisotropy_F2}:

\begin{equation}
    \text{K}_\mathrm{eff}=\frac{\mu_0\int_0^{M_s}[H^{\mathrm{eff}}(\theta;M)-H^{\mathrm{eff}}(0^\circ;M)]dM}{\sin^2\theta}
\end{equation}
\noindent where $M_s$ is the saturation magnetization (taken from the $\theta$=0$^\circ$ $M$ vs. $H$ curve). At $T$ = 2.5 K, for $\theta$=90$^\circ$ $\text{K}_\mathrm{eff}$ calculated to be 14.6$\times$10$^5$ J m$^{-3}$, which shows excellent agreement with the earlier report on a nearly similar composition~\cite{Cr1.53Te2_Adv}. The change of $\text{K}_\mathrm{eff}$ with intercalated Cr concentration has been shown in Fig.~\ref{Anisotropy}(b) [obtained from~\cite{Cr1.53Te2_Adv, Cr1+deltaTe2_PRM}]. The variation of $\text{K}_\mathrm{eff}$ with angle has been shown in the left panel of Fig.~\ref{Anisotropy}(c).




\par
The magnetic anisotropy energy is the energy required to rotate the magnetization from the easy axis to another direction (e.g., the hard axis). For a uniaxial system, with the easy axis perpendicular to the plane, the total anisotropy energy is often given by:

\begin{equation}
    \text{E}_{\mathrm{total}}=\text{K}_\mathrm{eff}.\sin^2\theta.\text{V}_{\mathrm{uc}}
\end{equation}

\noindent here $\text{V}_{\mathrm{uc}}$ is the unit cell volume. Taking $\theta$=90$^\circ$ (magnetization along the hard axis, in-plane) and $\text{V}_{\mathrm{uc}}$=80.88 \AA$^3$ (taken from diffraction study), the total anisotropy energy found to be approximately $\text{E}_{\mathrm{total}}$= 7.37$\times$10$^{-4}$ eV. As our unit cell has approximately 1.33 magnetic atoms (n$_{mag}$= 1.33 Cr), therefore the anisotropy energy per magnetic atom, $\text{E}_{\mathrm{MAE}}=\text{E}_{\mathrm{total}}$/n$_{mag}$= 5.54$\times$10$^{-4}$ eV = 554 $\mu$eV. The angle dependence of $\text{E}_{\mathrm{MAE}}$ has been shown in the right panel of Fig.~\ref{Anisotropy}(c) and it follows the theoretical curve very well [see Fig.5 (e) in the main text].

\vspace{0.5 cm}
\section{Theoretical study}
\label{Theoretical study_H}
\vspace{0.3 cm}

As discussed in our previous paper~\cite{PRM_Cr1.33Te2}, the supercell Cr$_{10}$Te$_{16}$ (equivalent to Cr$_{1.25}$Te$_2$) was used for the theoretical calculations [see Table~\ref{tab:Cr1.25Te2}]; therefore, the theoretical magnetic investigations were also carried out on this composition. The possible magnetic structures corresponding to this composition as obtain from the MAXMAGN program are listed in the table~\ref{table:MAXMAGN1}. The magnetic space groups $\textit{P}\overline{3}'m1$ and $\textit{P}\overline{3}m1$ are not permissible, as they constrain the moments to lie within the $ab$-plane. In case of $\textit{P}\overline{3}'m'1$, the Cr3 and Cr4 sites carry no magnetic moment. Therefore, the investigation proceeds with $\textit{P}\overline{3}m'1$, which has also been adopted for the experimental analysis. From table~\ref{table:MAXMAGN1} it is evident that Cr2 at the 6$i$ site has a canted magnetic moment.

\begin{table}[t]
\centering
\caption{\sffamily Crystallographic information Cr$_{10}$Te$_{16}$ (equivalent to Cr$_{1.25}$Te$_2$): Space Group: \textit{P}$\overline{3}$m1 (\#164)}
\label{tab:Cr1.25Te2}
\begin{tabular}{ccccccc}
\hline\hline
\textbf{Atom} & \textbf{Label} & \textbf{Site} & \textbf{Sym.} & $x$ & $y$ & $z$ \\
\hline
Cr & Cr1 & 2$c$ & 3m.  & 0.00000 & 0.00000 & 0.25000 \\
Cr & Cr2 & 6$i$ & .m.  & 0.00000 & 0.50000 & 0.25000 \\
Cr & Cr3 & 1$a$ & $-3$m. & 0.00000 & 0.00000 & 0.00000 \\
Cr & Cr4 & 1$b$ & $-3$m. & 0.00000 & 0.00000 & 0.50000 \\
Te & Te1 & 6$i$ & .m. & 0.33333 & 0.16667 & 0.12400 \\
Te & Te2 & 6$i$ & .m. & 0.33333 & 0.16667 & 0.62400 \\
Te & Te3 & 2$d$ & 3m. & 0.33333 & 0.66667 & 0.12400 \\
Te & Te4 & 2$d$ & 3m. & 0.33333 & 0.66667 & 0.62400 \\
\hline\hline
\end{tabular}

\vspace{6pt}
\begin{tabular}{ccccccc}
\textbf{Parameter} & $a$ (\AA) & $b$ (\AA) & $c$ (\AA) & $\alpha (^{\circ})$ & $\beta (^{\circ})$ & $\gamma (^{\circ})$ \\
\hline
Value & 7.85962 & 7.85962 & 12.11328 & 90.00 & 90.00 & 120.00 \\
\end{tabular}
\end{table}

\begin{table}[h]
\centering
\caption{\sffamily Symmetry-allowed magnetic moments for Cr sites in different magnetic space groups. Moment components are given as $(M_x, M_y, M_z)$.}
\label{table:MAXMAGN1}
\resizebox{0.8\columnwidth}{!}{%
\begin{tabular}{|c|c|c|c|c|}
\hline
   & \multicolumn{4}{|c|}{Allowed magnetic space groups} \\ 
\hline
 \shortstack{Magnetic \\ atoms} & \textit{P}$\overline{3}$m$'$1 & \textit{P}$\overline{3}'$m$'$1 & \textit{P}$\overline{3}'$m1 & \textit{P}$\overline{3}$m1 \\ 
 & (\#164.89) & (\#164.88) & (\#164.87) & (\#164.85) \\ 
\hline
Cr1 & (0, 0, $M_z$) & (0, 0, $M_z$) & (0, 0, 0) & (0, 0, 0) \\
Cr2 & ($M_x$, -$M_x$, $M_z$) & ($M_x$, -$M_x$, $M_z$) & ($M_x$, $M_x$, 0) & ($M_x$, $M_x$, 0) \\
Cr3 & (0, 0, $M_z$) & (0, 0, 0) & (0, 0, 0) & (0, 0, 0) \\
Cr4 & (0, 0, $M_z$) & (0, 0, 0) & (0, 0, 0) & (0, 0, 0) \\
\hline
\end{tabular}
}
\end{table}

\par
At first, systematic vacancies were introduced into the original cell (Cr$_{10}$Te$_{16}$) at both the Te and Cr sites, resulting in a final cell with the composition Cr$_{8}$Te$_{12}$ ($\approx$ Cr$_{1.33}$Te$_2$). Subsequently, a systematic investigation of the magnetic ground state was carried out for two distinct cases: the system without vacancies (WoV) and the system with vacancies (WV). 
\vspace{0.2 cm}
\section{Computational details}
\label{Computational details_I}
\vspace{0.2 cm}

The electronic structure calculations were performed within density functional theory (DFT) using the Vienna \textit{Ab-initio} Simulation Package (VASP)~\cite{hafner2008ab}. The interaction between ions and valence electrons was treated using the projector-augmented wave (PAW) method~\cite{kresse1999ultrasoft} with a plane-wave cutoff energy of 600~eV. Exchange--correlation effects were described within the generalized gradient approximation (GGA)~\cite{perdew}, supplemented by an on-site Coulomb term (GGA + U) to account for electron correlation in the localized orbitals. The effective Coulomb interaction parameter $U$ was varied between 0.5 and 0.8~eV to reproduce the experimental magnetic moments. Brillouin-zone integrations were performed using a Monkhorst--Pack $10\times10\times10$ $k$-point mesh. Structural relaxations were carried out until the forces on all atoms were below 0.001~eV/\AA{} and the total energy converged within $10^{-6}$~eV. 

\vspace{0.2 cm}
\section{Ground- and Excited-State Energy Landscapes of WoV and WV States}
\label{Ground- and Excited-State Energy Landscapes of WoV and WV States_J}
\vspace{0.2 cm}

\begin{figure}[ht]
\centering
\includegraphics[width = 15 cm]{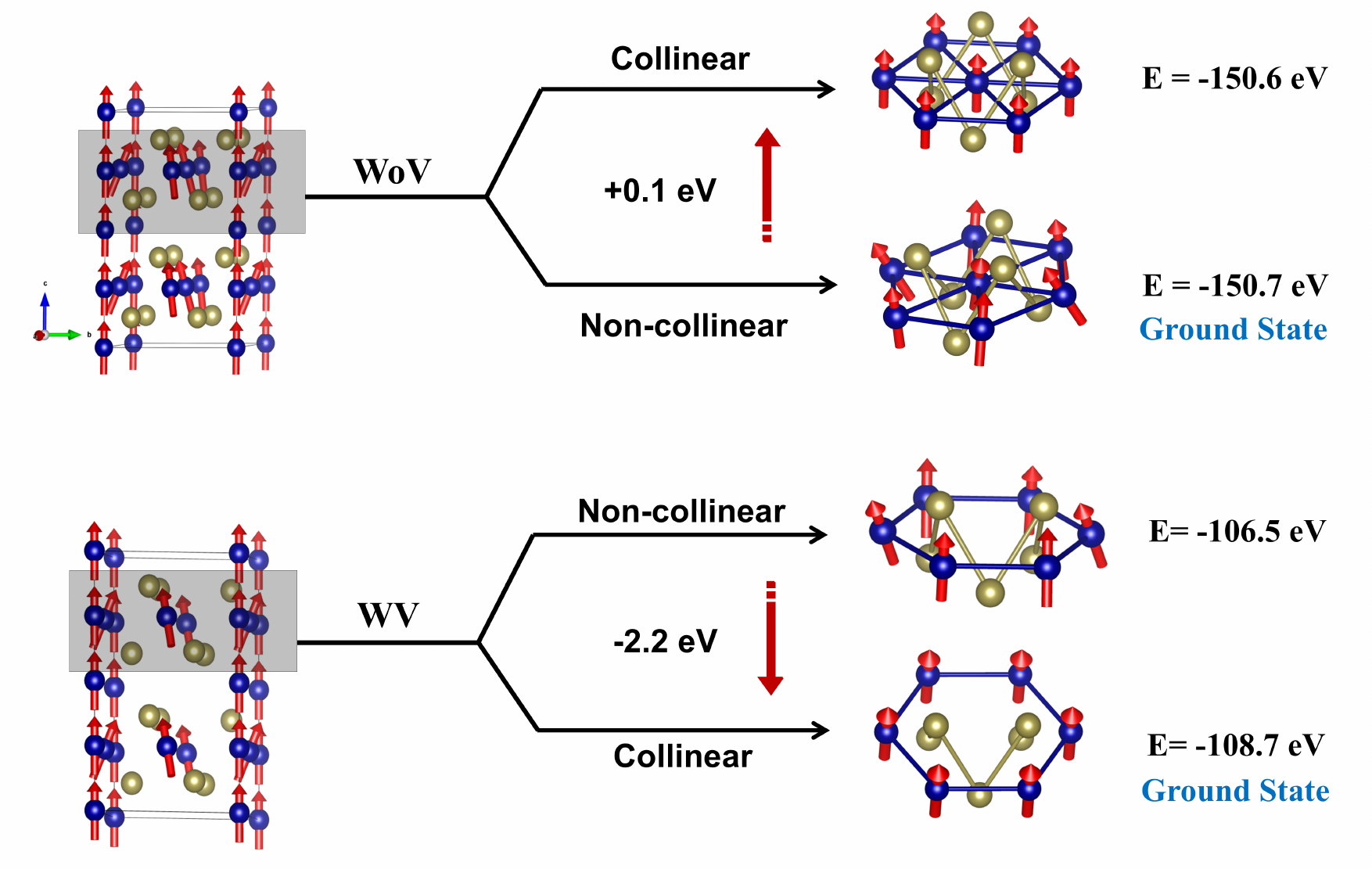}
\caption{\sffamily The upper panel shows the favorable magnetic configuration for the system without vacancy  \emph{i.e.,} Cr$_{10}$Te$_{16}$. The lower panel shows the favorable magnetic configuration in presence of vacancy \emph{i.e.,} for Cr$_{8}$Te$_{12}$.}
\label{Theory}
\end{figure}

For the without-vacancy (WoV) and with-vacancy (WV) states, both collinear and non-collinear spin configurations were considered at the 6$i$ site to examine the stability of magnetic states under spin-orbit coupling (SOC). In the collinear case, the spin quantization axis was fixed along the [0~0~1] direction, while in the non-collinear case, the magnetic moment at the 6$i$ site was oriented along [0.25~ -0.25~ 1.00], as shown in Fig.~\ref{Theory}.

\noindent The upper panel of Fig.~\ref{Theory} shows that, for the WoV case, the noncollinear configuration represents the true ground state of the system, whereas the lower panel indicates that, for the WV case, the collinear configuration is energetically favored.

\vspace{0.2 cm}
\section{Onsite energy extraction from Wannier functions}
\label{Onsite energy extraction from Wannier functions_K}
\vspace{0.2 cm}

Onsite energy corresponding to a specific atomic orbital, we performed Wannierization of the Bloch wavefunctions using the maximally localized Wannier functions (MLWFs) framework~\cite{pizzi2020wannier90,MarzariPhysRevB.56.12847}. Starting from the Bloch eigenstates $\psi_{n\mathbf{k}}(\mathbf{r})$ calculated within density functional theory (DFT), we constructed Wannier functions via a unitary transformation in reciprocal space,  

\begin{equation}
|w_{n\mathbf{R}}\rangle = \frac{V}{(2\pi)^3} \int_{\text{BZ}} d\mathbf{k}\, e^{-i\mathbf{k}\cdot\mathbf{R}} \sum_m U_{mn}^{(\mathbf{k})} |\psi_{m\mathbf{k}}\rangle ,
\end{equation}

where $U_{mn}^{(\mathbf{k})}$ is a gauge choice optimized to minimize the spread functional. The resulting Wannier Hamiltonian in real space is obtained as  

\begin{equation}
H_{mn}(\mathbf{R}) = \langle w_{m\mathbf{0}} | \hat{H} | w_{n\mathbf{R}} \rangle .
\end{equation}

The onsite energy ($\epsilon_{i}$) of a particular orbital localized on atom $i$ is then identified as the diagonal matrix element of the Wannier Hamiltonian at the home cell ($\mathbf{R}=0$),  

\begin{equation}
\epsilon_{i} = H_{ii}(\mathbf{R}=0).
\end{equation}
 
 The Bloch functions were constructed within the maximally localized Wannier function (MLWF) framework, using atomic-like Cr-$d$ orbitals as initial projectors for 6i site's Cr in both cases.

\begin{table}[h]
\centering
\begin{tabular}{c|c|c}
\hline
\textbf{Orbital} & \textbf{$\epsilon_{i}$(eV) WoV Case} & \textbf{$\epsilon_{i}$(eV) WV Case} \\
\hline
$xy$          & 4.6  & 5.2  \\
$yz$          & 5.3  & 4.95 \\
$xz$          & 4.6  & 5.15 \\
$z^{2}$       & 5.5  & 5.35 \\
$x^{2}-y^{2}$ & 5.5  & 5.5  \\
\hline
\end{tabular}

\begin{picture}(0,0)
  \put(-135,48){$\left\{\rule{0pt}{20pt}\right.$}
  \put(-150,48){$t_{2g}$}

  \put(-135,12){$\left\{\rule{0pt}{12pt}\right.$}
  \put(-150,12){$e_{g}$}
\end{picture}

\caption{Onsite energies (in eV) for WoV and WV cases for different $d$ orbitals of Cr.}
\label{table:onsite_energy}
\end{table}

\vspace{0.2 cm}
\section{Octahedral geometry}
\label{Octahedral geometry_L}
\vspace{0.2 cm}

\begin{figure}[ht]
\centering
\includegraphics[width = 14 cm]{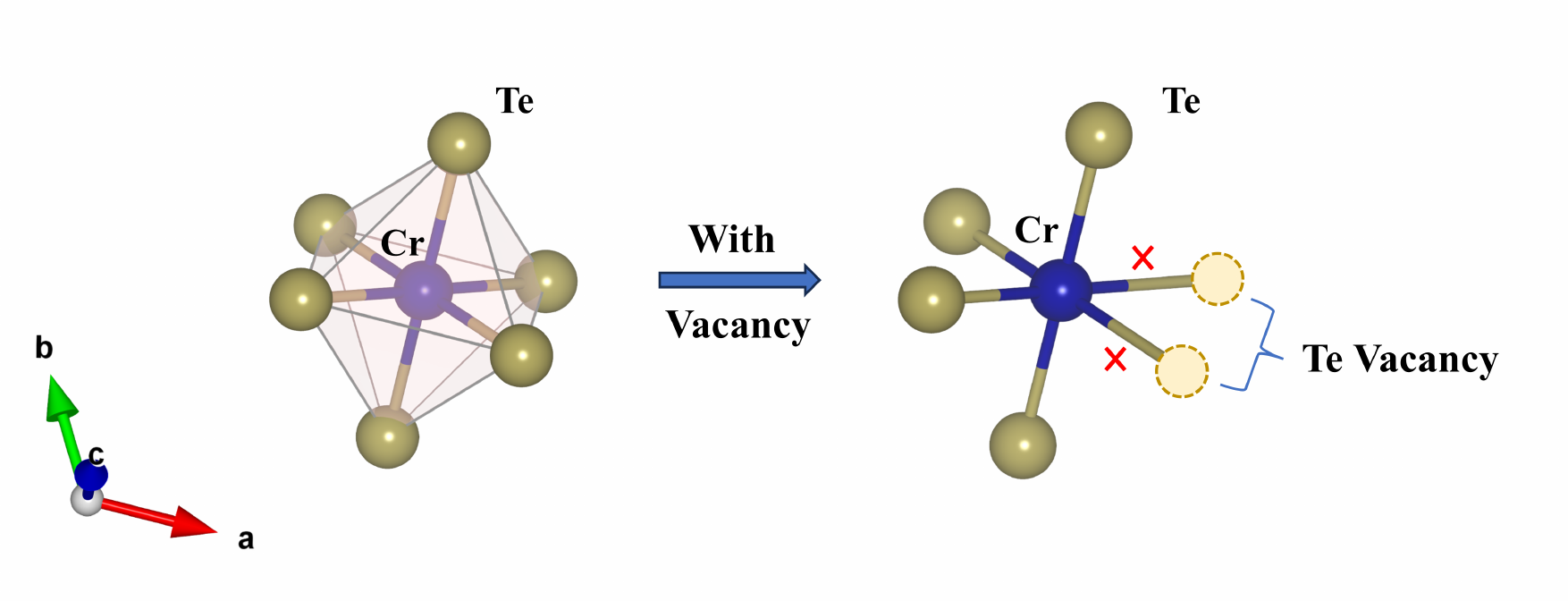}
\caption{\sffamily Octahedral environment of Cr without and with vacancies}
\label{Octahedra}
\end{figure}

In Fig.~\ref{Octahedra} the octahedral geometry of Cr atom has been displayed before and after the introduction of vacancies. Here we specially focuses on the Cr atom at the 6$i$ site, which has a canted magnetic moment.

\vspace{0.2 cm}
\section{Perturbative Spin-Orbit Coupling Formalism for MAE}
\label{Perturbative Spin-Orbit Coupling Formalism for MAE_M}
\vspace{0.2 cm}

The magnetocrystalline anisotropy energy (MAE) was evaluated within second-order perturbation theory of the spin-orbit coupling (SOC) Hamiltonian~\cite{PhysRevB.84.195430,PhysRevB.95.165415}. For an atom with orbital angular momentum $\mathbf{L}$ and spin $\mathbf{S}$, the SOC Hamiltonian takes the form
\begin{equation}
H_{\mathrm{SOC}} = \xi \, \mathbf{L}\cdot \mathbf{S} 
= \xi \left( L_x S_x + L_y S_y + L_z S_z \right),
\end{equation}
where $\xi$ is the SOC strength, $L_{x,y,z}$ are the orbital angular momentum matrices in the $d$-orbital basis ($l=2$), and $S_{x,y,z}$ are the spin-$\tfrac{1}{2}$ operators.

The orbital angular momentum operators are constructed from the ladder operators,
\begin{align}
L_+ |m\rangle &= \sqrt{l(l+1) - m(m+1)} \, |m+1\rangle, \\
L_- |m\rangle &= \sqrt{l(l+1) - m(m-1)} \, |m-1\rangle,
\end{align}
with
\begin{equation}
L_x = \tfrac{1}{2}(L_+ + L_-), \qquad
L_y = -\tfrac{i}{2}(L_+ - L_-), \qquad
L_z = \mathrm{diag}(m),
\end{equation}
for $m=-2,-1,0,1,2$.

For a general magnetization axis $\hat{n}$, the SOC Hamiltonian is rotated via
\begin{equation}
H_{\mathrm{SOC}}(\hat{n}) = R \, H_{\mathrm{SOC}}(z)\, R^\dagger,
\end{equation}
where $R = I_{\mathrm{orb}} \otimes U$, with $U$ the $2\times 2$ spin rotation matrix
\begin{equation}
U = \exp\left[-i \, \theta \, (\hat{r}\cdot \mathbf{S}) \right],
\end{equation}
determined by the rotation aligning $z \to \hat{n}$.

The second-order SOC energy correction is then expressed as
\begin{equation}
E^{(2)}(\hat{n}) = - \sum_{i \in \mathrm{occ}} \sum_{j \in \mathrm{unocc}}
\frac{ \big| \langle \psi_j | H_{\mathrm{SOC}}(\hat{n}) | \psi_i \rangle \big|^2 }{\varepsilon_j - \varepsilon_i},
\end{equation}
where $|\psi_i\rangle$ and $|\psi_j\rangle$ are unperturbed $d$-orbital $\otimes$ spin states with corresponding energies $\varepsilon_i$ and $\varepsilon_j$, and the summations run over occupied ($\mathrm{occ}$) and unoccupied ($\mathrm{unocc}$) states, as defined by the input occupations along the global $z$-axis.

\vspace{0.2cm}
\section{Hamiltonian Matrix Representation Before and After Spin–Orbit Coupling}
\label{Hamiltonian Matrix Representation Before and After Spin–Orbit Coupling_N}
\vspace{0.2cm}
\subsection{WoV system}
We construct the tight-binding Hamiltonian in the five $d$-orbital basis
\[
\{\, d_{xy}\uparrow,\; d_{yz}\uparrow,\; d_{z^2}\uparrow,\; d_{xz}\uparrow,\; d_{x^2-y^2}\uparrow,\;
   d_{xy}\downarrow,\; d_{yz}\downarrow,\; d_{z^2}\downarrow,\; d_{xz}\downarrow,\; d_{x^2-y^2}\downarrow \,\},
\]
where the first five states correspond to spin-up and the remaining five to spin-down components.

To analyze the influence of spin–orbit coupling (SOC) on the low-energy electronic structure, we constructed the model Hamiltonian both before and after including SOC effects. The non–SOC Hamiltonian ($H_0$) is purely diagonal, representing the orbital-resolved on-site energies, while the inclusion of SOC introduces small off-diagonal elements that couple different spin and orbital states.

The Hamiltonian before including SOC is given by:

\begin{equation}
H_0 =
\begin{pmatrix}
4.6 & 0 & 0 & 0 & 0 & 0 & 0 & 0 & 0 & 0 \\
0 & 5.3 & 0 & 0 & 0 & 0 & 0 & 0 & 0 & 0 \\
0 & 0 & 5.5 & 0 & 0 & 0 & 0 & 0 & 0 & 0 \\
0 & 0 & 0 & 4.6 & 0 & 0 & 0 & 0 & 0 & 0 \\
0 & 0 & 0 & 0 & 5.5 & 0 & 0 & 0 & 0 & 0 \\
0 & 0 & 0 & 0 & 0 & 4.6 & 0 & 0 & 0 & 0 \\
0 & 0 & 0 & 0 & 0 & 0 & 5.3 & 0 & 0 & 0 \\
0 & 0 & 0 & 0 & 0 & 0 & 0 & 5.5 & 0 & 0 \\
0 & 0 & 0 & 0 & 0 & 0 & 0 & 0 & 4.6 & 0 \\
0 & 0 & 0 & 0 & 0 & 0 & 0 & 0 & 0 & 5.5
\end{pmatrix}
\end{equation}

After including the SOC interaction, the Hamiltonian becomes:
\begin{equation}
H = H_0 + H_{\mathrm{SOC}} =
\begin{pmatrix}
4.55 & 0 & 0 & 0.05 & 0 & 0 & 0 & 0 & 0 & 0 \\
0 & 5.35 & 0 & 0 & 0 & 0 & 0 & 0 & 0 & 0 \\
0 & 0 & 5.475 & 0 & 0 & 0.0612 & 0 & 0 & 0 & 0 \\
0.05 & 0 & 0 & 4.625 & 0 & 0 & 0 & 0 & 0 & 0 \\
0 & 0 & 0 & 0 & 5.5 & 0 & 0 & 0.0612 & 0 & 0 \\
0 & 0 & 0.0612 & 0 & 0 & 4.6 & 0 & 0 & 0 & 0 \\
0 & 0 & 0 & 0 & 0 & 0 & 5.325 & 0 & 0 & 0.05 \\
0 & 0 & 0 & 0 & 0.0612 & 0 & 0 & 5.475 & 0 & 0 \\
0 & 0 & 0 & 0 & 0 & 0 & 0 & 0 & 4.65 & 0 \\
0 & 0 & 0 & 0 & 0 & 0 & 0.05 & 0 & 0 & 5.45
\end{pmatrix}
\end{equation}

Here, the diagonal elements correspond to the orbital energy levels, while the small off-diagonal terms arise due to SOC-induced mixing between different orbital and spin components.

\vspace{0.3cm}
\subsection{WV system}
\vspace{0.2 cm}
\subsection*{Hamiltonian before SOC (\(H_0\))}

In the absence of spin-orbit coupling (SOC), the Hamiltonian is diagonal, containing only the on-site energies of the $d$ orbitals. The diagonal terms represent the crystal-field-split energies of individual orbitals, while the absence of any off-diagonal elements reflects no hybridization between orbitals in this simplified model. The corresponding Hamiltonian matrix is

The Hamiltonian before including SOC is given by:

\begin{equation}
H_0 =
\begin{pmatrix}
5.2 & 0 & 0 & 0 & 0 & 0 & 0 & 0 & 0 & 0 \\
0 & 4.95 & 0 & 0 & 0 & 0 & 0 & 0 & 0 & 0 \\
0 & 0 & 5.15 & 0 & 0 & 0 & 0 & 0 & 0 & 0 \\
0 & 0 & 0 & 5.35 & 0 & 0 & 0 & 0 & 0 & 0 \\
0 & 0 & 0 & 0 & 5.5 & 0 & 0 & 0 & 0 & 0 \\
0 & 0 & 0 & 0 & 0 & 5.2 & 0 & 0 & 0 & 0 \\
0 & 0 & 0 & 0 & 0 & 0 & 4.95 & 0 & 0 & 0 \\
0 & 0 & 0 & 0 & 0 & 0 & 0 & 5.15 & 0 & 0 \\
0 & 0 & 0 & 0 & 0 & 0 & 0 & 0 & 5.35 & 0 \\
0 & 0 & 0 & 0 & 0 & 0 & 0 & 0 & 0 & 5.5
\end{pmatrix}
\end{equation}

After including the SOC interaction, the Hamiltonian becomes:

\begin{equation}
H = H_0 + H_{\mathrm{SOC}} =
\begin{pmatrix}
5.15 & 0 & 0 & 0.05 & 0 & 0 & 0 & 0 & 0 & 0 \\
0 & 5.00 & 0 & 0 & 0 & 0 & 0 & 0 & 0 & 0 \\
0 & 0 & 5.125 & 0 & 0 & 0.0612 & 0 & 0 & 0 & 0 \\
0.05 & 0 & 0 & 5.375 & 0 & 0 & 0 & 0 & 0 & 0 \\
0 & 0 & 0 & 0 & 5.5 & 0 & 0 & 0.0612 & 0 & 0 \\
0 & 0 & 0.0612 & 0 & 0 & 5.2 & 0 & 0 & 0 & 0 \\
0 & 0 & 0 & 0 & 0 & 0 & 4.975 & 0 & 0 & 0.05 \\
0 & 0 & 0 & 0 & 0.0612 & 0 & 0 & 5.125 & 0 & 0 \\
0 & 0 & 0 & 0 & 0 & 0 & 0 & 0 & 5.4 & 0 \\
0 & 0 & 0 & 0 & 0 & 0 & 0.05 & 0 & 0 & 5.45
\end{pmatrix}
\end{equation}

Here, the diagonal elements correspond to the orbital energy levels, while the small off-diagonal terms arise due to SOC-induced mixing between different orbital and spin components.

Here, the diagonal elements slightly shift compared to $H_0$, reflecting SOC-induced renormalization of orbital energies, while the off-diagonal elements (typically $\sim 0.05$–$0.06$~eV) correspond to SOC-mediated mixing between orbitals such as $(d_{xy}, d_{xz})$ and $(d_{z^2}, d_{xy})$. These hybridization terms are essential in generating MAE.

\vspace{0.2cm}
\section{MAE Characteristics of the WoV System}
\label{MAE Characteristics of the WoV System_O}
\vspace{0.3 cm}

\begin{figure}[H]
\centering
\includegraphics[width = 16 cm]{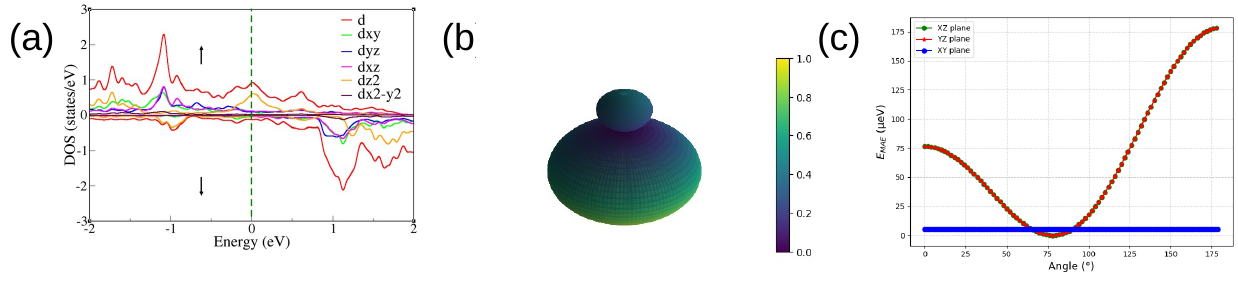}
\caption{\sffamily (a) Partial density of states (PDOS) of Cr $d$ orbitals for spin-up and spin-down channels. (b) Normalized magnetocrystalline anisotropy energy (MAE) on sphere. (c) MAE in different planes as a function of the polar angle.}
\label{MAE_WoV}
\end{figure}

In the WoV system, our density functional theory (DFT) calculations reveal that the ground state adopts a ferromagnetically noncollinear spin configuration. Taking this configuration as the magnetic ground state, we computed the partial density of states (PDOS), as displayed in Fig.~\ref{MAE_WoV}(a). The normalized magnetic anisotropy energy (MAE), presented in Fig.~\ref{MAE_WoV}(b), demonstrates that the $z$ axis is no longer the easy axis of magnetization. This conclusion is further supported by Fig.~\ref{MAE_WoV}(c), which shows a finite MAE in the $xz$ and $yz$ planes at $0^\circ$, confirming the deviation from z.

\bibliography{Reference}

@article{Cr0.87Te_ACS_Nano,
author = {Liu, Jun and Ding, Bei and Liang, Jinjing and Li, Xue and Yao, Yuan and Wang, Wenhong},
title = {Magnetic Skyrmionic Bubbles at Room Temperature and Sign Reversal of the Topological Hall Effect in a Layered Ferromagnet {Cr$_{0.87}$Te}},
journal = {ACS Nano},
volume = {16},
number = {9},
pages = {13911-13918},
year = {2022},
doi = {10.1021/acsnano.2c02928},
    note ={PMID: 36000915},

URL = { 
    
        https://doi.org/10.1021/acsnano.2c02928
    
},

}

@Article{Cr1.33Te2_Rana_Saha,
author={Saha, Rana
and Meyerheim, Holger L.
and G{\"o}bel, B{\"o}rge
and Hazra, Binoy Krishna
and Deniz, Hakan
and Mohseni, Katayoon
and Antonov, Victor
and Ernst, Arthur
and Knyazev, Dmitry
and Bedoya-Pinto, Amilcar
and Mertig, Ingrid
and Parkin, Stuart S. P.},
title={Observation of N{\'e}el-type skyrmions in acentric self-intercalated {Cr$_{1+\delta}$Te$_2$}},
journal={Nature Communications},
year={2022},
month={Jul},
day={08},
volume={13},
number={1},
pages={3965},
issn={2041-1723},
doi={10.1038/s41467-022-31319-y},
url={https://doi.org/10.1038/s41467-022-31319-y}
}

@article{Cr1.53Te2_Adv,
author = {Zhang, Chenhui and Liu, Chen and Zhang, Junwei and Yuan, Youyou and Wen, Yan and Li, Yan and Zheng, Dongxing and Zhang, Qiang and Hou, Zhipeng and Yin, Gen and Liu, Kai and Peng, Yong and Zhang, Xi-Xiang},
title = {Room-Temperature Magnetic Skyrmions and Large Topological Hall Effect in Chromium Telluride Engineered by Self-Intercalation},
journal = {Advanced Materials},
volume = {35},
number = {1},
pages = {2205967},
doi = {https://doi.org/10.1002/adma.202205967},
url = {https://advanced.onlinelibrary.wiley.com/doi/abs/10.1002/adma.202205967},
year = {2023}
}

@article{Cr5Te8_Nitesh,
author = {Rai, Banik and Kuila, Sandip Kumar and Saha, Rana and Hazra, Sankalpa and De, Chandan and Sau, Jyotirmoy and Gopalan, Venkatraman and Jana, Partha Pratim and Parkin, Stuart S. P. and Kumar, Nitesh},
title = {Peculiar Magnetic and Magneto-Transport Properties in a Noncentrosymmetric Self-Intercalated van der Waals Ferromagnet {Cr$_5$Te$_8$}},
journal = {Chemistry of Materials},
volume = {37},
number = {2},
pages = {746-755},
year = {2025},
doi = {10.1021/acs.chemmater.4c02996},

URL = { 
    
        https://doi.org/10.1021/acs.chemmater.4c02996
    
    
},

}

@article{CrTe2_AdvF,
author = {Liu, Ying and Liu, Yangrui and Dan, Jiadong and Liu, Wei and Wang, Luyang and Hu, Kejun and Wang, Weiwei and Zhang, Lei and Ge, Binghui and Du, Haifeng and Song, Dongsheng},
title = {Atomic-Scale Order and Disorder Induced Diverse Topological Spin Textures in Self-Intercalated Van der Waals Magnets $\mathrm{Cr_{1+\delta}Te_2}$
},
journal = {Advanced Functional Materials},
volume = {35},
number = {5},
pages = {2414699},
doi = {https://doi.org/10.1002/adfm.202414699},
url = {https://advanced.onlinelibrary.wiley.com/doi/abs/10.1002/adfm.202414699},
year = {2025}
}

@article{Cr5Te8_AHE,
  title = {Anomalous Hall effect in the trigonal {$Cr_{5}Te_{8}$} single crystal},
  author = {Liu, Yu and Petrovic, C.},
  journal = {Phys. Rev. B},
  volume = {98},
  issue = {19},
  pages = {195122},
  numpages = {5},
  year = {2018},
  month = {Nov},
  publisher = {American Physical Society},
  doi = {10.1103/PhysRevB.98.195122},
  url = {https://link.aps.org/doi/10.1103/PhysRevB.98.195122}
}

@article{Cr-Te_Phase,
title = {Transition metal-chalcogen systems viii: The {Cr-Te} phase diagram},
journal = {Journal of the Less Common Metals},
volume = {92},
number = {2},
pages = {265-282},
year = {1983},
issn = {0022-5088},
doi = {https://doi.org/10.1016/0022-5088(83)90493-9},
url = {https://www.sciencedirect.com/science/article/pii/0022508883904939},
author = {Herbert Ipser and Kurt L. Komarek and Kurt O. Klepp}
}

@article{Fujisawa_Cr1+deltaTe2,
author = {Fujisawa, Yuita and Pardo-Almanza, Markel and Hsu, Chia-Hsiu and Mohamed, Atwa and Yamagami, Kohei and Krishnadas, Anjana and Chang, Guoqing and Chuang, Feng-Chuan and Khoo, Khoong Hong and Zang, Jiadong and Soumyanarayanan, Anjan and Okada, Yoshinori},
title = {Widely Tunable Berry Curvature in the Magnetic Semimetal {$Cr_{1+\delta}Te_{2}$}},
journal = {Advanced Materials},
volume = {35},
number = {12},
pages = {2207121},
doi = {https://doi.org/10.1002/adma.202207121},
url = {https://advanced.onlinelibrary.wiley.com/doi/abs/10.1002/adma.202207121},
year = {2023}
}

@article{Cr1+deltaTe2_PRM,
  title = {Tailoring magnetism in self-intercalated {$Cr_{1+\delta}Te_{2}$} epitaxial films},
  author = {Fujisawa, Y. and Pardo-Almanza, M. and Garland, J. and Yamagami, K. and Zhu, X. and Chen, X. and Araki, K. and Takeda, T. and Kobayashi, M. and Takeda, Y. and Hsu, C. H. and Chuang, F. C. and Laskowski, R. and Khoo, K. H. and Soumyanarayanan, A. and Okada, Y.},
  journal = {Phys. Rev. Mater.},
  volume = {4},
  issue = {11},
  pages = {114001},
  numpages = {7},
  year = {2020},
  month = {Nov},
  publisher = {American Physical Society},
  doi = {10.1103/PhysRevMaterials.4.114001},
  url = {https://link.aps.org/doi/10.1103/PhysRevMaterials.4.114001}
}

@article{Cr1.2Te2_Nano_Letter,
author = {Huang, Meng and Gao, Lei and Zhang, Ying and Lei, Xunyong and Hu, Guojing and Xiang, Junxiang and Zeng, Hualing and Fu, Xuewen and Zhang, Zengming and Chai, Guozhi and Peng, Yong and Lu, Yalin and Du, Haifeng and Chen, Gong and Zang, Jiadong and Xiang, Bin},
title = {Possible Topological Hall Effect above Room Temperature in Layered $\mathrm{Cr_{1.2}Te_2}$ Ferromagnet},
journal = {Nano Letters},
volume = {21},
number = {10},
pages = {4280-4286},
year = {2021},
doi = {10.1021/acs.nanolett.1c00493},
    note ={PMID: 33979154},

URL = { 
    
        https://doi.org/10.1021/acs.nanolett.1c00493
},


}

@article{Cr1.61Te2_THE,
author = {Huang, Yalei and Zuo, Na and Zhang, Zheyi and Xing, Xiangzhuo and Yao, Xinyu and Zhang, Anlei and Ma, Haowei and Xu, Chunqiang and Jiao, Wenhe and Zhou, Wei and Sankar, Raman and Qian, Dong and Xu, Xiaofeng},
title = {In-Plane Magnetic Anisotropy and Large Topological Hall Effect in Self-Intercalated Ferromagnet {Cr$_{1.61}$Te$_2$}},
journal = {Advanced Functional Materials},
volume = {n/a},
number = {n/a},
pages = {e10351},
year = {2025},
doi = {https://doi.org/10.1002/adfm.202510351},
url = {https://advanced.onlinelibrary.wiley.com/doi/abs/10.1002/adfm.202510351}
}

@article{Cr2.76Te4,
  title = {Investigation of the anomalous and topological Hall effects in layered monoclinic ferromagnet {Cr$_{2.76}$Te$_4$}},
  author = {Purwar, Shubham and Low, Achintya and Bose, Anumita and Narayan, Awadhesh and Thirupathaiah, S.},
  journal = {Phys. Rev. Mater.},
  volume = {7},
  issue = {9},
  pages = {094204},
  numpages = {9},
  year = {2023},
  month = {Sep},
  publisher = {American Physical Society},
  doi = {10.1103/PhysRevMaterials.7.094204},
  url = {https://link.aps.org/doi/10.1103/PhysRevMaterials.7.094204}
}

@article{Cr1.22Te2,
  title = {Evidence for magnetoelastic coupling and chiral magnetic ground state in quasi-van der Waals trigonal {Cr$_{1.22}$Te$_2$}},
  author = {Hossain, S. M. and Rai, B. and Baral, P. R. and Zaharko, O. and Kumar, N. and Bera, A. K. and Majumder, M.},
  journal = {Phys. Rev. B},
  volume = {112},
  issue = {2},
  pages = {024439},
  numpages = {12},
  year = {2025},
  month = {Jul},
  publisher = {American Physical Society},
  doi = {10.1103/qnx5-ckyr},
  url = {https://link.aps.org/doi/10.1103/qnx5-ckyr}
}

@article{Cr5Te8_nonCo,
title = {A neutron diffraction study of structural and magnetic properties of monoclinic {Cr$_5$Te$_8$}},
journal = {Solid State Sciences},
volume = {10},
number = {8},
pages = {1099-1105},
year = {2008},
issn = {1293-2558},
doi = {https://doi.org/10.1016/j.solidstatesciences.2007.11.013},
url = {https://www.sciencedirect.com/science/article/pii/S1293255807003573},
author = {Zhong-Le Huang and Winfried Kockelmann and Mark Telling and Wolfgang Bensch}
}

@article{Cr2Te3_nonCo,
title = {Neutron diffraction study of {Cr$_2$Te$_3$} single crystal},
journal = {Solid State Communications},
volume = {16},
number = {7},
pages = {895-897},
year = {1975},
issn = {0038-1098},
doi = {https://doi.org/10.1016/0038-1098(75)90888-1},
url = {https://www.sciencedirect.com/science/article/pii/0038109875908881},
author = {T. Hamasaki and T. Hashimoto and Y. Yamaguchi and H. Watanabe}
}

@article{Cr3Te4_Cr5Te6_nonCo,
  title={The Magnetic Structure of {Cr$_2$Te$_3$}, {Cr$_3$Te$_4$}, and {Cr$_5$Te$_6$.}},
  author={Arne F. Andresen and Eila Zeppezauer and T. Boive and Bo Nordstr{\"o}m and C I Br{\"a}nd{\'e}n},
  journal={Acta Chemica Scandinavica},
  year={1970},
  volume={24},
  pages={3495-3509},
  url={https://api.semanticscholar.org/CorpusID:95772121}
}

@article{Cr2Te3_Cr5Te6_collinear,
  author    = {Arne F. Andresen},
  title     = {A Neutron Diffraction Investigation of {Cr$_{2}$Te$_{3}$} and {Cr$_{5}$Te$_{6}$}},
  journal   = {Acta Chemica Scandinavica},
  volume    = {17},
  year      = {1963},
  pages     = {1335--1342},
  publisher = {Scandinavian Chemical Society},
  address   = {Kjeller, Norway},
  url       = {https://scispace.com/pdf/a-neutron-diffraction-investigation-of-cr-sub-2-te-sub-3-and-42lesn5kym.pdf}
}

@article{PRM_Cr1.33Te2,
  title = {Suppression of intrinsic Hall effect through competing Berry curvature in {Cr$_{1+\delta}$Te$_2$}},
  author = {Chowdhury, Prasanta and Sau, Jyotirmay and Numan, Mohamad and Sannigrahi, Jhuma and Gutmann, Matthias and Giri, Saurav and Kumar, Manoranjan and Majumdar, Subham},
  journal = {Phys. Rev. Mater.},
  volume = {9},
  issue = {2},
  pages = {024407},
  numpages = {12},
  year = {2025},
  month = {Feb},
  publisher = {American Physical Society},
  doi = {10.1103/PhysRevMaterials.9.024407},
  url = {https://link.aps.org/doi/10.1103/PhysRevMaterials.9.024407}
}

@article{SXD,
author = "Keen, David A. and Gutmann, Matthias J. and Wilson, Chick C.",
title = "{SXD {--} the single-crystal diffractometer at the ISIS spallation neutron source}",
journal = "Journal of Applied Crystallography",
year = "2006",
volume = "39",
number = "5",
pages = "714--722",
month = "Oct",
doi = {10.1107/S0021889806025921},
url = {https://doi.org/10.1107/S0021889806025921},
}

@article{SXD2001,
  author       = {Gutmann, M.},
  title        = {SXD2001 -- A program for treating data from TOF neutron single-crystal diffraction},
  journal      = {Acta Crystallographica Section A},
  year         = {2005},
  volume       = {61},
  pages        = {c164},
  doi          = {10.1107/S0108767305093025},
}

@article{Jana2020,
url = {https://doi.org/10.1515/zkri-2023-0005},
title = {Jana2020– a new version of the crystallographic computing system Jana},
author = {Václav Petříček and Lukáš Palatinus and Jakub Plášil and Michal Dušek},
pages = {271--282},
volume = {238},
number = {7-8},
journal = {Zeitschrift für Kristallographie - Crystalline Materials},
doi = {doi:10.1515/zkri-2023-0005},
year = {2023},
lastchecked = {2025-03-20}
}

@article{Cr12-xTe16,
  title = {Structure, chromium vacancies, and magnetism in a $\mathrm{C}{\mathrm{r}}_{12\ensuremath{-}x}\mathrm{T}{\mathrm{e}}_{16}$ compound},
  author = {Cao, Guixin and Zhang, Qiang and Frontzek, Matthias and Xie, Weiwei and Gong, Dongliang and Sterbinsky, George E. and Jin, Rongying},
  journal = {Phys. Rev. Mater.},
  volume = {3},
  issue = {12},
  pages = {125001},
  numpages = {8},
  year = {2019},
  month = {Dec},
  publisher = {American Physical Society},
  doi = {10.1103/PhysRevMaterials.3.125001},
  url = {https://link.aps.org/doi/10.1103/PhysRevMaterials.3.125001}
}

@ARTICLE{Bilbao,
	author = {Aroyo, M.I. and Perez-Mato, J.M. and Orobengoa, D. and Tasci, E. and De La Flor, G. and Kirov, A.},
	title = {Crystallography online: Bilbao crystallographic server},
	year = {2011},
	journal = {Bulgarian Chemical Communications},
	volume = {43},
	number = {2},
	pages = {183 – 197},
	url = {https://www.scopus.com/inward/record.uri?eid=2-s2.0-80955140447&partnerID=40&md5=488772b9e21d2636a3952f66ae80ae84},
	type = {Article},
	publication_stage = {Final},
	source = {Scopus},
}

@article{MAXMGN,
   author = "Perez-Mato, J.M. and Gallego, S.V. and Tasci, E.S. and Elcoro, L. and de la Flor, G. and Aroyo, M.I.",
   title = "Symmetry-Based Computational Tools for Magnetic Crystallography", 
   journal= "Annual Review of Materials Research",
   year = "2015",
   volume = "45",
   number = "Volume 45, 2015",
   pages = "217-248",
   doi = "https://doi.org/10.1146/annurev-matsci-070214-021008",
   url = "https://www.annualreviews.org/content/journals/10.1146/annurev-matsci-070214-021008",
   publisher = "Annual Reviews",
   issn = "1545-4118",
   type = "Journal Article",
  }

@article{Elettra,
doi = {10.1088/1742-6596/190/1/012043},
url = {https://dx.doi.org/10.1088/1742-6596/190/1/012043},
year = {2009},
month = {nov},
publisher = {},
volume = {190},
number = {1},
pages = {012043},
author = {Andrea Di Cicco and Giuliana Aquilanti and Marco Minicucci and Emiliano Principi and Nicola Novello and Andrea Cognigni and Luca Olivi},
title = {Novel XAFS capabilities at ELETTRA synchrotron light source},
journal = {Journal of Physics: Conference Series}
}

@article{EXAFS1,
author = "Ravel, B. and Newville, M.",
title = "{{\it ATHENA}, {\it ARTEMIS}, {\it HEPHAESTUS}: data analysis for X-ray absorption spectroscopy using {\it IFEFFIT}}",
journal = "Journal of Synchrotron Radiation",
year = "2005",
volume = "12",
number = "4",
pages = "537--541",
month = "Jul",
doi = {10.1107/S0909049505012719},
url = {https://doi.org/10.1107/S0909049505012719},
}

@article{EXAFS2,
author = "Newville, Matthew",
title = "{{\it IFEFFIT}: interactive XAFS analysis and {\it FEFF} fitting}",
journal = "Journal of Synchrotron Radiation",
year = "2001",
volume = "8",
number = "2",
pages = "322--324",
month = "Mar",
doi = {10.1107/S0909049500016964},
url = {https://doi.org/10.1107/S0909049500016964},
}

@article{FEFF6,
author = "Newville, Matthew",
title = "{EXAFS analysis using {\it FEFF} and {\it FEFFIT}}",
journal = "Journal of Synchrotron Radiation",
year = "2001",
volume = "8",
number = "2",
pages = "96--100",
month = "Mar",
doi = {10.1107/S0909049500016290},
url = {https://doi.org/10.1107/S0909049500016290},
}

@book{Judd1963,
  title     = {Operator Techniques in Atomic Spectroscopy},
  author    = {B. R. Judd},
  publisher = {McGraw--Hill},
  year      = {1963}
}

@article{Cooper1979,
  title   = {Theory of spin-orbit-induced magnetic anisotropy},
  author  = {Cooper, B. R. and Siemann, R. and Cromer, D. T. and Andersen, O. K.},
  journal = {Phys. Rev. B},
  volume  = {19},
  number  = {3},
  pages   = {1317--1327},
  year    = {1979},
  publisher = {American Physical Society},
  doi     = {10.1103/PhysRevB.19.1317}
}

@article{Bruno1989,
  title   = {Tight-binding approach to the orbital magnetic moment and magnetocrystalline anisotropy of transition-metal monolayers},
  author  = {Bruno, P.},
  journal = {Phys. Rev. B},
  volume  = {39},
  number  = {1},
  pages   = {865--868},
  year    = {1989},
  publisher = {American Physical Society},
  doi     = {10.1103/PhysRevB.39.865}
}

@article{song2022superexchange,
  title = {Superexchange and spin-orbit coupling in monolayer and bilayer chromium trihalides},
  author = {Song, Kok Wee and Fal'ko, Vladimir I.},
  journal = {Phys. Rev. B},
  volume = {106},
  issue = {24},
  pages = {245111},
  numpages = {16},
  year = {2022},
  month = {Dec},
  publisher = {American Physical Society},
  doi = {10.1103/PhysRevB.106.245111},
  url = {https://link.aps.org/doi/10.1103/PhysRevB.106.245111}
}

@article{koseki2019spin,
author = {Koseki, Shiro and Matsunaga, Nikita and Asada, Toshio and Schmidt, Michael W. and Gordon, Mark S.},
title = {Spin–Orbit Coupling Constants in Atoms and Ions of Transition Elements: Comparison of Effective Core Potentials, Model Core Potentials, and All-Electron Methods},
journal = {The Journal of Physical Chemistry A},
volume = {123},
number = {12},
pages = {2325-2339},
year = {2019},
doi = {10.1021/acs.jpca.8b09218},
    note ={PMID: 30817150},

URL = { 
    
        https://doi.org/10.1021/acs.jpca.8b09218
    
    

},

}

@article{PhysRevB.100.144413,
  title = {Theoretical investigation of magnetic anisotropy at the $\mathrm{L}{\mathrm{a}}_{0.5}\mathrm{S}{\mathrm{r}}_{0.5}\mathrm{Mn}{\mathrm{O}}_{3}/\mathrm{LaCo}{\mathrm{O}}_{2.5}$ interface},
  author = {Chen, Xiaobing and Zhang, Shihao and Liu, Banggui and Hu, Fengxia and Shen, Baogen and Sun, Jirong},
  journal = {Phys. Rev. B},
  volume = {100},
  issue = {14},
  pages = {144413},
  numpages = {8},
  year = {2019},
  month = {Oct},
  publisher = {American Physical Society},
  doi = {10.1103/PhysRevB.100.144413},
  url = {https://link.aps.org/doi/10.1103/PhysRevB.100.144413}
}

@article{PhysRevResearch.4.013237,
  title = {Effect of the valence state on the band magnetocrystalline anisotropy in two-dimensional rare-earth/noble-metal compounds},
  author = {Blanco-Rey, M. and Castrillo-Bodero, R. and Ali, K. and Gargiani, P. and Bertran, F. and Sheverdyaeva, P. M. and Ortega, J. E. and Fernandez, L. and Schiller, F.},
  journal = {Phys. Rev. Res.},
  volume = {4},
  issue = {1},
  pages = {013237},
  numpages = {9},
  year = {2022},
  month = {Mar},
  publisher = {American Physical Society},
  doi = {10.1103/PhysRevResearch.4.013237},
  url = {https://link.aps.org/doi/10.1103/PhysRevResearch.4.013237}
}

@article{PhysRevB.81.104426,
  title = {Magnetocrystalline anisotropy energy of Co and Fe adatoms on the (111) surfaces of Pd and Rh},
  author = {B\l{}o\ifmmode \acute{n}\else \'{n}\fi{}ski, Piotr and Lehnert, Anne and Dennler, Samuel and Rusponi, Stefano and Etzkorn, Markus and Moulas, G\'eraud and Bencok, Peter and Gambardella, Pietro and Brune, Harald and Hafner, J\"urgen},
  journal = {Phys. Rev. B},
  volume = {81},
  issue = {10},
  pages = {104426},
  numpages = {18},
  year = {2010},
  month = {Mar},
  publisher = {American Physical Society},
  doi = {10.1103/PhysRevB.81.104426},
  url = {https://link.aps.org/doi/10.1103/PhysRevB.81.104426}
}

@misc{Cr1deltaTe2_neutron_2025,
  title        = {{ISIS} Neutron and Muon Facility},
  author       = {Jhuma Sannigrahi and Prasanta Chowdhury and Devashi Adroja and Matthias Gutmann and Subham Majumdar},
  year         = {2025},
  doi          = {10.5286/ISIS.E.RB2410336},
  url          = {https://doi.org/10.5286/ISIS.E.RB2410336}
}

@misc{SI,
  title = {Orbital Degeneracy Control through Vacancies: A Pathway to Tailored Magnetic State in {Cr$_{1+\delta}$Te$_2$}},
  author = {Chowdhury, Prasanta and Sau, Jyotirmoy and Numan, Mohamad and Sannigrahi, Jhuma and Gutmann, Matthias and Das, Gangadhar and Giri, Saurav and Kumar, Manoranjan and Majumdar, Subham},
  year = {2025},
  note = {{\textbf{Supplemental Material}}},
}

@article{Anisotropy_F1,
  title = {Tunable Carr-type temperature dependence of uniaxial magnetocrystalline anisotropy in Fe-deficient ${\mathrm{Fe}}_{3\ensuremath{-}x}{\mathrm{GeTe}}_{2}$},
  author = {Liu, Jiawei and Zhou, Liang and Liang, Rui and Li, Shuilin and Li, Ziying and Zhang, Li and Li, Zishuang and Wang, Haozhe and Huang, Yi and Liu, Ronghua and Tang, Nujiang},
  journal = {Phys. Rev. B},
  volume = {111},
  issue = {14},
  pages = {144425},
  numpages = {12},
  year = {2025},
  month = {Apr},
  publisher = {American Physical Society},
  doi = {10.1103/PhysRevB.111.144425},
  url = {https://link.aps.org/doi/10.1103/PhysRevB.111.144425}
}

@article{Anisotropy_F2,
author = {Sucksmith, Willie  and Thompson, J. E. },
title = {The magnetic anisotropy of cobalt},
journal = {Proceedings of the Royal Society of London. Series A. Mathematical and Physical Sciences},
volume = {225},
number = {1162},
pages = {362-375},
year = {1954},
doi = {10.1098/rspa.1954.0209},

URL = {https://royalsocietypublishing.org/doi/abs/10.1098/rspa.1954.0209}
}

@article{hafner2008ab,
author = {Hafner, Jürgen},
title = {Ab-initio simulations of materials using VASP: Density-functional theory and beyond},
journal = {Journal of Computational Chemistry},
volume = {29},
number = {13},
pages = {2044-2078},
doi = {https://doi.org/10.1002/jcc.21057},
url = {https://onlinelibrary.wiley.com/doi/abs/10.1002/jcc.21057},

year = {2008}
}

@article{perdew,
  title = {Accurate and simple analytic representation of the electron-gas correlation energy},
  author = {Perdew, John P. and Wang, Yue},
  journal = {Phys. Rev. B},
  volume = {45},
  issue = {23},
  pages = {13244--13249},
  numpages = {0},
  year = {1992},
  month = {Jun},
  publisher = {American Physical Society},
  doi = {10.1103/PhysRevB.45.13244},
  url = {https://link.aps.org/doi/10.1103/PhysRevB.45.13244}
}

@article{pizzi2020wannier90,
doi = {10.1088/1361-648X/ab51ff},
url = {https://dx.doi.org/10.1088/1361-648X/ab51ff},
year = {2020},
month = {jan},
publisher = {IOP Publishing},
volume = {32},
number = {16},
pages = {165902},
author = {Giovanni Pizzi and Valerio Vitale and Ryotaro Arita and others},
title = {Wannier90 as a community code: new features and applications},
journal = {Journal of Physics: Condensed Matter},

}

@article{MarzariPhysRevB.56.12847,
  title = {Maximally localized generalized Wannier functions for composite energy bands},
  author = {Marzari, Nicola and Vanderbilt, David},
  journal = {Phys. Rev. B},
  volume = {56},
  issue = {20},
  pages = {12847--12865},
  numpages = {0},
  year = {1997},
  month = {Nov},
  publisher = {American Physical Society},
  doi = {10.1103/PhysRevB.56.12847},
  url = {https://link.aps.org/doi/10.1103/PhysRevB.56.12847}
}

@article{kresse1999ultrasoft,
  title={From ultrasoft pseudopotentials to the projector augmented-wave method},
  author={Kresse, Georg and Joubert, Daniel},
  journal={Physical review b},
  volume={59},
  number={3},
  pages={1758},
  year={1999},
  publisher={APS}
}

@article{PhysRevB.84.195430,
  title = {Low-energy effective Hamiltonian involving spin-orbit coupling in silicene and two-dimensional germanium and tin},
  author = {Liu, Cheng-Cheng and Jiang, Hua and Yao, Yugui},
  journal = {Phys. Rev. B},
  volume = {84},
  issue = {19},
  pages = {195430},
  numpages = {11},
  year = {2011},
  month = {Nov},
  publisher = {American Physical Society},
  doi = {10.1103/PhysRevB.84.195430},
  url = {https://link.aps.org/doi/10.1103/PhysRevB.84.195430}
}

@article{PhysRevB.95.165415,
  title = {Model spin-orbit coupling Hamiltonians for graphene systems},
  author = {Kochan, Denis and Irmer, Susanne and Fabian, Jaroslav},
  journal = {Phys. Rev. B},
  volume = {95},
  issue = {16},
  pages = {165415},
  numpages = {19},
  year = {2017},
  month = {Apr},
  publisher = {American Physical Society},
  doi = {10.1103/PhysRevB.95.165415},
  url = {https://link.aps.org/doi/10.1103/PhysRevB.95.165415}
}

@article{Cr2Te3_Theory_GS,
  title = {Evolution of ground state in ${\mathrm{Cr}}_{2}{\mathrm{Te}}_{3}$ single crystal under applied magnetic field},
  author = {Jiang, Z. Z. and Liang, X. and Luo, X. and Gao, J. J. and Wang, W. and Wang, T. Y. and Yang, X. C. and Wang, X. L. and Zhang, L. and Sun, Y. and Tong, P. and Hu, J. F. and Song, W. H. and Lu, W. J. and Sun, Y. P.},
  journal = {Phys. Rev. B},
  volume = {106},
  issue = {9},
  pages = {094407},
  numpages = {11},
  year = {2022},
  month = {Sep},
  publisher = {American Physical Society},
  doi = {10.1103/PhysRevB.106.094407},
  url = {https://link.aps.org/doi/10.1103/PhysRevB.106.094407}
}

@Article{CrTe2_Theory,
author ="Feng, Dushuo and Shen, Zhong and Xue, Yufei and Guan, Zhihao and Xiao, Runhu and Song, Changsheng",
title  ="Strain-induced magnetic phase transition{,} magnetic anisotropy switching and bilayer antiferromagnetic skyrmions in van der Waals magnet CrTe2",
journal  ="Nanoscale",
year  ="2023",
volume  ="15",
issue  ="4",
pages  ="1561-1567",
publisher  ="The Royal Society of Chemistry",
doi  ="10.1039/D2NR04740C",
url  ="http://dx.doi.org/10.1039/D2NR04740C"}

@article{CrTe_Theory,
doi = {10.1088/0953-8984/22/15/156002},
url = {https://doi.org/10.1088/0953-8984/22/15/156002},
year = {2010},
month = {mar},
publisher = {},
volume = {22},
number = {15},
pages = {156002},
author = {Polesya, S and Mankovsky, S and Benea, D and Ebert, H and Bensch, W},
title = {Finite-temperature magnetism of CrTe and CrSe},
journal = {Journal of Physics: Condensed Matter}
}

@article{Cr2Te3_Unconvetional_AHE,
author = {He, Keke and Bian, Mengying and Seddon, Samuel D. and Jagadish, Koushik and Mucchietto, Andrea and Ren, He and Kirstein, Erik and Asadi, Reza and Bai, Jaeil and Yao, Chao and Pan, Sheng and Yu, Jie-Xiang and Milde, Peter and Huai, Chang and Hui, Haolei and Zang, Jiadong and Sabirianov, Renat and Cheng, Xuemei M. and Miao, Guoxing and Xing, Hui and Shao, Yu-Tsun and Crooker, Scott A. and Eng, Lukas and Hou, Yanglong and Bird, Jonathan P. and Zeng, Hao},
title = {Unconventional Anomalous Hall Effect Driven by Self-Intercalation in Covalent 2D Magnet {Cr$_2$Te$_3$}},
journal = {Advanced Science},
volume = {12},
number = {2},
pages = {2407625},
doi = {https://doi.org/10.1002/advs.202407625},
url = {https://advanced.onlinelibrary.wiley.com/doi/abs/10.1002/advs.202407625},

}

@article{MnCoGe,
    author = {Wang, Jian-Tao and Wang, Ding-Sheng and Chen, Changfeng and Nashima, O. and Kanomata, T. and Mizuseki, H. and Kawazoe, Y.},
    title = {Vacancy-Induced Structural and Magnetic Transition in {MnCo$_{1-x}$Ge}},
    journal = {Applied Physics Letters},
    volume = {89},
    number = {26},
    pages = {262504},
    year = {2006},
    month = {12},
    issn = {0003-6951},
    doi = {10.1063/1.2424273},
    url = {https://doi.org/10.1063/1.2424273},
}

@article{BaFeSe,
  title = {Effect of iron vacancies on magnetic order and spin dynamics of the spin ladder {BaFe$_{2-\delta}$S$_{1.5}$Se$_{1.5}$}},
  author = {Liu, Zengjia and Ni, Xiao-Sheng and Li, Lisi and Sun, Hualei and Liang, Feixiang and Frandsen, Benjamin A. and Christianson, Andrew D. and dela Cruz, Clarina and Xu, Zhijun and Yao, Dao-Xin and Lynn, Jeffrey W. and Birgeneau, Robert J. and Cao, Kun and Wang, Meng},
  journal = {Phys. Rev. B},
  volume = {105},
  issue = {21},
  pages = {214303},
  numpages = {7},
  year = {2022},
  month = {Jun},
  publisher = {American Physical Society},
  doi = {10.1103/PhysRevB.105.214303},
  url = {https://link.aps.org/doi/10.1103/PhysRevB.105.214303}
}

@article{LaMnSb2,
  title = {Vacancy-tuned magnetism in {LaMn$_x$Sb$_2$}},
  author = {Slade, Tyler J. and Sapkota, Aashish and Wilde, John M. and Zhang, Qiang and Wang, Lin-Lin and Lapidus, Saul H. and Schmidt, Juan and Heitmann, Thomas and Bud'ko, Sergey L. and Canfield, Paul C.},
  journal = {Phys. Rev. Mater.},
  volume = {7},
  issue = {11},
  pages = {114203},
  numpages = {25},
  year = {2023},
  month = {Nov},
  publisher = {American Physical Society},
  doi = {10.1103/PhysRevMaterials.7.114203},
  url = {https://link.aps.org/doi/10.1103/PhysRevMaterials.7.114203}
}

\end{document}